\documentclass[aps,prd,10pt,twocolumn,superscriptaddress,showpacs,longbibliography,nofootinbib,noeprint,floatfix]{revtex4-2}

\pdfoutput=1
\usepackage{graphicx}
\usepackage{amsfonts,amsmath,amssymb,bm,bbm}
\usepackage{color}
\usepackage{enumitem}
\usepackage{url}  
\usepackage{ulem}

\usepackage[colorlinks=true,linkcolor=blue,citecolor=blue]{hyperref}
\usepackage[capitalize]{cleveref}
\usepackage[dvipsnames, usenames]{xcolor}
\usepackage{orcidlink}

\usepackage{graphicx}
\newcommand{\orcid}[1]{\href{https://orcid.org/#1}{\includegraphics[width=10pt]{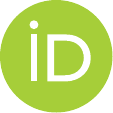}}}
\newcommand{\github}[1]{\href{https://github.com/#1}{\includegraphics[width=10pt]{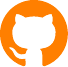}}}

\definecolor{romared}{RGB}{142,0,28}

\def\bi{\begin{itemize}[noitemsep,leftmargin=*]
\setlength\itemsep{1em}
        }
\def\ei{\end{itemize}}

\usepackage{lipsum}

\begin{document}


\title{Nonlinear aspects of stochastic particle acceleration}

\author{Martin Lemoine \orcid{0000-0002-2395-7812}}
\email{lemoine@iap.fr}
\affiliation{Institut d'Astrophysique de Paris, CNRS -- Sorbonne Université, Paris, France}

\author{Kohta Murase \orcid{0000-0002-5358-5642}}
\email{murase@psu.edu}
\affiliation{Department of Physics, The Pennsylvania State University, University Park, Pennsylvania 16802, USA}
\affiliation{Department of Astronomy and Astrophysics, The Pennsylvania State University, University Park, Pennsylvania 16802, USA}
\affiliation{Center for multi-messenger Astrophysics, Institute for Gravitation and the Cosmos, The Pennsylvania State University, University Park, Pennsylvania 16802, USA}
\affiliation{School of Natural Sciences, Institute for Advanced Study, Princeton, New Jersey 08540, USA}
\affiliation{Center for Gravitational Physics and Quantum Information, Yukawa Institute for Theoretical Physics, Kyoto, Kyoto 606-8502 Japan}

\author{Frank Rieger \orcid{0000-0003-1334-2993}}
\email{frank.rieger@ipp.mpg.de}
\affiliation{Max Planck Institute for Plasma Physics (IPP), Boltzmannstra{\ss}e 2,
85748 Garching, Germany}
\affiliation{Institute for Theoretical Physics, Heidelberg University, Philosophenweg 12, 69120 Heidelberg, Germany}

\date{\today}

\begin{abstract}
In turbulent magnetized plasmas, charged particles can be accelerated to high energies through their interactions with the turbulent motions. As they do so, they draw energy from the turbulence, possibly up to the point where they start modifying the turbulent cascade. Stochastic acceleration then enters a nonlinear regime because turbulence damping back-reacts in turn on the acceleration process. This article develops a phenomenological model to examine this situation in detail and to explore its consequences for the particle and turbulent energy spectra. We determine a criterion that specifies the threshold of nonthermal particle energy density and the characteristic momentum beyond which back-reaction becomes effective. Once the back-reaction sets in, the turbulence cascade becomes damped below a length scale that keeps increasing in time. The accelerated particle momentum distribution develops a near power-law of the form ${\rm d}n/{\rm d}p\propto p^{-s}$ with $s\sim2$ beyond the momentum at which back-reaction first sets in. At very high energies, where the gyroradius of accelerated particles becomes comparable to the outer scale of the turbulence, the energy spectrum can display an even harder spectrum with $s\sim 1.3-1.5$ over a short segment. The low-energy part of the spectrum, below the critical momentum, is expected to be hard ($s\sim 1$ or harder), and shaped by any residual acceleration process in the damped region of the turbulence cascade. This characteristic broken power-law shape with $s\sim 2$ at high energies may find phenomenological applications in various high-energy astrophysical contexts. 
\end{abstract}

\maketitle

\section{Introduction}
Particle acceleration in turbulent flows has long been recognized as a key physical process in multi-messenger astrophysics, notably thanks to its ability to dissipate turbulent energy into nonthermal populations and to accelerate particles up to high energies. Ever since its introduction by E. Fermi in his seminal paper on the origin of cosmic rays~\cite{1949PhRv...75.1169F}, stochastic particle acceleration has been invoked to model the origin of high-energy particles and radiation in extreme environments as diverse as the Galactic center~\cite{04Liu,2006ApJ...647.1099L,2011PhRvL.107i1101M}, accretion flows~\cite{Dermer:1995ju,Becker:2009dh,14Lynn,2015ApJ...806..159K,Kimura:2016fjx,2019MNRAS.485..163K,2021NatCo..12.5615K} and black hole coronae~\cite{2020PhRvL.125a1101M,2021ApJ...922...45K,2022ApJ...941L..17M,2023arXiv230111327G}, extragalactic jets~\cite{09Sullivan,2011ApJ...739...66T,2016ApJ...816...24K,2017ApJ...842...39L,18Asano,2022MNRAS.517L..16T,2022MNRAS.517.2502S}, intracluster gas~\cite{2007MNRAS.378..245B,2008ApJ...682..175P,2011MNRAS.410..127B,2016MNRAS.458.2584B,17Eckert}, gamma-ray bursts~\cite{1996ApJ...461L..37B,2016PhRvD..94b3005A,2017ApJ...846L..28X,2021PhRvD.104j3005Z}, pulsar wind nebulae~\cite{2016JPlPh..82d6301L,2017ApJ...841...78T,2019MNRAS.489.2403L,2023ApJ...953..116L} etc.

For the purpose of phenomenological applications, the nonthermal energy spectra are commonly derived from the solution of a Fokker-Planck transport equation characterized by a momentum-dependent diffusion coefficient $D_{pp}$. Often implicit to that formulation is a test-particle picture, in which the turbulence physics is set by the physical conditions at the driving scale on the one hand, while nonthermal particles independently draw energy from the turbulent flow on the other. Provided the rate at which energy is absorbed by the particles remains small compared to that which flows through the cascade from large to small scales, this test-particle picture may indeed suffice. Yet, as realized early on~\cite{1979ApJ...229..413E,1979ApJ...230..373E}, continuous stirring of the turbulence keeps injecting energy into the nonthermal particles, possibly to the point where they start damping the cascade. This issue is particularly relevant for turbulent acceleration, which is known to shape hard momentum ($p$) distributions in the time-asymptotic limit, of the form ${\rm d}n/{\rm d}p\propto p^{-s}$ with $s \sim 1$~\cite{1984A&A...136..227S,2006ApJ...647..539B,2008ApJ...681.1725S}. 

Once back-reaction becomes significant, stochastic particle acceleration becomes nonlinear, in the sense that altering the turbulent cascade feeds back on the acceleration process, hence on the redistribution of energy among the nonthermal particles, which itself governs the back-reaction. The ensuing phenomenology has been addressed in the frame of a quasilinear Fokker-Planck model in the context of solar flares~\cite{1979ApJ...229..413E,1995ApJ...452..912M}, pulsar wind nebulae~\cite{2023ApJ...953..116L}, clusters of galaxies~\cite{2007MNRAS.378..245B} or extragalactic jets~\cite{2016ApJ...816...24K,2021PhRvD.104j3005Z,2022MNRAS.517.2502S}. The aim of the present work is to examine in detail the evolution of the coupled dynamical system and to determine the ensuing particle energy distribution, adopting a more general and more agnostic perspective than these studies: more general, because we do not have a specific application in mind and correspondingly set aside additional effects such as energy losses and particle escape to focus on the generic spectral shape (although we eventually discuss the influence of such effects); more agnostic, because we intend to remain as inclusive as possible regarding the nature of the coupling between particles and turbulence.

We find that in the regime of strong damping, the nonthermal particle energy spectrum takes on a generic broken power-law shape, with a hard slope at low energies ($s\lesssim 1$) and flat segment at high energies ($s\sim 2$). Interestingly, such spectra are rather common features of high-energy multi-messenger phenomenology. The origin of that spectral shape can be briefly described as follows. Once the particle energy density crosses a threshold, which we determine, turbulence damping becomes effective below a scale $r_{\rm g}(p_{\rm pk})$, where $p_{\rm pk}$ denotes the momentum at which the particle energy spectrum peaks at that time, and $r_{\rm g}(p_{\rm pk})$ the associated gyroradius. The momentum $p_{\rm pk}$ also corresponds to the break of the final broken power-law spectrum. Once turbulent damping sets in, the low-energy segment freezes because of the lack of turbulent power on the relevant scales. Meanwhile, damping continues to occur on length scales $\lesssim r_{\rm g}(p_{\rm d})$, where $p_{\rm d}$, and hence $r_{\rm g}(p_{\rm d})$ increases with time. This evolution shapes the high-energy segment with constant energy per decade. In the text, we provide a more refined description of that spectral shape as a function of time and discuss possible variations in the presence of energy losses, escape as well as potential secondary acceleration scenarios. 

Our study remains exploratory in nature, given that many aspects of stochastic acceleration and turbulence physics remain poorly understood. Nevertheless, we find that the above results appear reasonably robust. The phenomenology also depends on some external constraints inherent to the source of turbulence, such as the timescale over which turbulence is driven, which we assume here to be large compared to all other scales, as well as the characteristic velocity of the eddies, which sets the acceleration timescale. In the relativistic regime (characteristic eddy velocity $\sim c$), back-reaction could become significant even if the driving time becomes as short as a few eddy turn-over times on the outer scale.

The discussion is organized as follows: we specify the physical model and its assumptions in Sec.~\ref{sec:model}, then discuss the implications for the particle spectra in Sec.~\ref{sec:impl}, and finally discuss those results in Sec.~\ref{sec:disc}. Throughout this work, we use $Q_x/Q=10^{x}$ in CGS units and set units such that $c=k_{\rm B}=1$.

\section{Physical model}\label{sec:model}
We consider a generic model in which a nonthermal population of relativistic particles is subject to a stochastic Fermi process. The turbulent plasma is assumed to be homogeneous (at least, statistically speaking) throughout space as well as steady in time. This means, in particular, that even though energy is continuously injected to drive the turbulence at the outer scale $\ell_{\rm c}$, the statistical properties governing the turbulence, {\it e.g.} the Alfvén velocity $v_{\rm A}$ (in terms of the background, coherent magnetic field $B_0$) and the relative amplitude of turbulent fluctuations at the outer scale $\delta_B \equiv \delta B/B$, remain constant. For simplicity, we also assume a magnetized regime with beta parameter $\beta \lesssim 1$. Additionally, throughout the main text, we assume that the driving of the turbulence takes place over timescales that are long compared to the other relevant timescales, in particular the eddy turn-over time on the outer scale, $t_{\rm nl}\sim \ell_{\rm c}/v_{\rm A}$, and the acceleration timescales $t_{\rm acc}$. We will discuss how the situation depicted here can be generalized to other situations in Sec.~\ref{sec:disc}.

\subsection{Particle acceleration in turbulence}\label{sec:linear}
In rather general terms, stochastic acceleration can lead to advection and diffusion in momentum space. Advection itself can result from net energy gain in the turbulence due to some asymmetry between energy-gain and loss processes, or from the momentum dependence of the diffusion coefficient in a purely diffusive process (the noise induced drift of stochastic processes). Advection plays a crucial role for what concerns turbulent damping, because it generically dominates the increase of the (nonthermal) particle energy density. 

To quantify the evolution in time of the differential particle energy spectrum $\mathcal E_p(t)\equiv 4\pi p^4 f(p,\,t) \equiv p^2\,{\rm d}n/{\rm d}p$, where $f(p,\,t)$ represents the particle distribution function and $n$ the particle density, we introduce two models, with, as a first option, the common diffusive Fokker-Planck model characterized by
\begin{equation}
    \partial_t f = \frac{1}{p^2}\partial_p\left(D_{pp}\,p^2\,\partial_p\, f\right),\,\label{eq:model1}
\end{equation}
characterized by the diffusion coefficient $D_{pp}$, for which we assume the generic form $D_{pp} = \nu_p p^2$ with $\nu_p$ a momentum-independent frequency, in agreement with recent measurements carried out in numerical particle-in-cell (PIC) experiments at large turbulence amplitude $\delta_B \gtrsim 1$ and in the large Alfv\'enic velocity regime ($v_{\rm A}\gtrsim 0.1\,c$)~\cite{17Zhdankin,2018ApJ...867L..18Z,18Comisso,2019ApJ...886..122C,2020ApJ...893L...7W,2021ApJ...922..172Z,2021ApJ...921...87N}. This hard-sphere diffusion model has become standard in phenomenological models of high-energy astrophysical sources based on stochastic acceleration. 

It has been realized however that the Fokker-Planck model cannot account for the accelerated particle spectra found in these numerical experiments, unless one adds an advection term with a non-trivial momentum dependence~\cite{2020ApJ...893L...7W,2021ApJ...922..172Z}. The numerical PIC results actually suggest that stochastic particle acceleration acts inhomogeneously throughout space, meaning that different particles experiencing vastly different energization schemes, in opposition to the common Brownian motion underlying the Fokker-Planck picture~\cite{2017PhRvL.119d5101I,2020MNRAS.499.4972L,2021ApJ...921...87N}. In this case, a more general transport equation appears necessary to describe the evolution in phase space. The scenario developed in Refs.~\cite{PhysRevD.104.063020,Bresci+22,2022PhRvL.129u5101L} argues in particular that the physical process at play can be seen as a generalization of the original Fermi process, in which particles gain most of their energy through localized interactions with intense, intermittent structures laid on scales of the order of the particle gyroradius or larger. This model reproduces satisfactorily the power-law spectra observed in those simulations, albeit admittedly at the price of increased complexity. To mimic the phenomenological features of that model, while retaining a level of simplicity adapted to the present discussion, we introduce a second scenario in which we initialize the spectrum as a (broken) power-law distribution and consider only the effect of advection, namely
\begin{equation}
    \partial_t f = -\frac{1}{p^2}\partial_p\left(A_p\,p^2 f\,\right),\,\label{eq:model2}
\end{equation}
where $A_p$ characterizes advection in momentum space. We assume a scaling $A_p = \nu_pp$, with $\nu_p$ a momentum-independent frequency, such as introduced for the Fokker-Planck model. This equation preserves the original power-law shape and propagates it to larger momenta in a way similar to the noise-induced drift of the diffusive model. In the following, it will be referred to as the ``power-law model''.

\begin{figure}[h]
\includegraphics[width=0.48\textwidth]{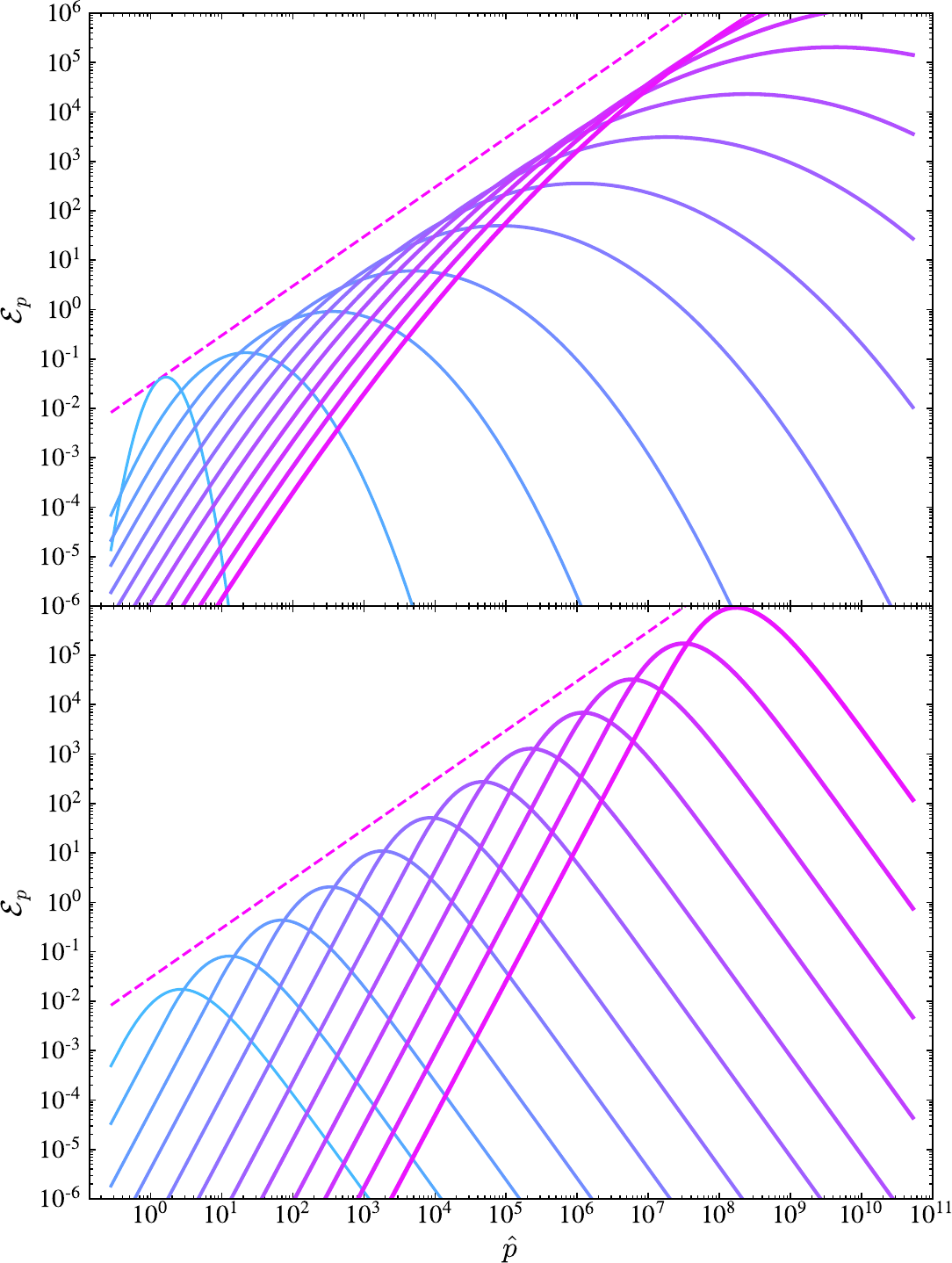}
\caption{Time-dependent solutions for the differential particle energy density spectrum per log-momentum interval $\mathcal E_p(t)\equiv 4\pi p^4 f(p,t)$, for impulsive injection of particles at momentum $p_0$ at time $t=0$, plotted as a function of $\hat p\equiv p/p_0$. Top panel: solution for the Fokker-Planck model described by Eq.~(\ref{eq:model1}); bottom panel: solution of the power-law model described by Eq.~(\ref{eq:model2}) with an initial index $s_{\rm hi}=4$ at high energies, $s_{\rm li}=-2$ at low energies (see text for detail). Lines plotted in light-thin to deep-thick colors range from early to late times and are plotted at constant time intervals $\Delta t$ such that $\nu_p\,\Delta t=1.0$ (resp. $1.6$) for model 1 (resp. model 2). The dashed line indicates a scaling $\mathcal E_p\propto p$. Turbulence damping by particle acceleration has not been taken into account here.
\label{fig:nofdbck}}
\end{figure}

The solutions to the above transport equations in the absence of back-reaction on the turbulence are shown in Figure~\ref{fig:nofdbck} for illustrative purposes: the upper panel shows the evolution in model 1 (``Fokker-Planck''), while the lower panel corresponds to model 2 (``power-law''). Lines are plotted from light-thin to deep-thick colors at constant time intervals from early to late. 
Model 1 assumes the impulsive injection of a density $n$ at the initial time, distributed as a narrow Maxwellian distribution in momentum, ${\rm d}n/{\rm d}p\propto p^2\,\exp\left[-(p-p_0)^2/(2\sigma_p^2)\right]$. Here, $p_0$ denotes a reference momentum and $\sigma_p\simeq 0.1$ an ad hoc initial width. 
In model 2, we adopt a broken power-law, ${\rm d}n/{\rm d}p \propto p^{-s_{\rm li}}\,\left(1 + p/p_0\right)^{-s_{\rm hi}+s_{\rm li}}$. 
We initialize the power-law spectrum with $s_{\rm li}=-2$, corresponding to the relativistic Maxwellian, and we adopt $s_{\rm hi}=4$ as observed in the sub-relativistic limit in magnetohydrodynamic (MHD) and PIC simulations evincing turbulent acceleration~\cite{Bresci+22,2022PhRvL.129u5101L,2022ApJ...936L..27C}. 
The dashed lines indicate a scaling $\mathcal E_p\propto p$ and provide the solution that would be obtained for continuous injection of particles at momentum $p_0$ at all times. 

Multiplying Eq.~(\ref{eq:model1}) or (\ref{eq:model2}) by $p$ and integrating by parts, it is easy to see that for $\nu_p\propto p^0$, the mean momentum $\langle p\rangle \equiv \int{\rm d}p\,p\,{\rm d}n/{\rm d}p$ increases exponentially in time at rate $4\nu_p$ (resp. $\nu_p$) for model 1 (resp. model 2). For continuous injection of particles at all times, one can check that the steady state solution of $\mathcal E_p$ at momenta $p\ll\langle p\rangle$ reads $\mathcal E_p\propto p/\nu_p$ in either model, see {\it e.g.}~\cite{2017ApJ...842...39L}. For $\nu_p\propto p^0$, one thus recover $\mathcal E_p\propto p$, as indicated by the dashed lines in Fig.~\ref{fig:nofdbck}. The scaling $D_{pp}\propto p^{5/3}$ based on quasilinear models is occasionally adopted~\cite{1984A&A...136..227S}; in such a case, $\mathcal E_p\propto p^{4/3}$. In all reasonable cases, therefore, $\mathcal E_p$ follows the behavior announced in the introduction: as acceleration proceeds, energy is fed to the accelerated particle population and its energy density increases with the peak energy of the time-dependent spectrum. On long timescales, back-reaction appears inevitable.

\subsection{Turbulence cascade and damping}
We now address the energy transfer between turbulent magnetic energy and the accelerated particles. The differential magnetic energy density spectrum of the turbulence cascade per log-wavenumber interval is written ${\mathcal E_B}_k$ and defined as ${\mathcal E_B}_k \equiv k {\mathcal S_B}_k$, in terms of the power spectrum of magnetic fluctuations ${\mathcal S_B}_k\propto k^{-5/3}$ (in the absence of feedback), with the normalization $\int{\rm d}\ln k\,{\mathcal E_B}_k \equiv \delta_B^2 \mathcal E_{B_0}$, where $\mathcal E_{B_0}\equiv B_0^2/(8\pi)$ denotes the magnetic energy density in the background magnetic field.  The turbulence anisotropy is implicitly contained in the full tri-dimensional dependence of the power spectrum of $\mathbf{k}$, but otherwise discarded here. For the sake of definiteness, we have assumed a Kolmogorov-type exponent $5/3$; our results can be trivially extended to an index $3/2$. In the absence of feedback, the energy spectrum ${\mathcal E_B}_k$ per log-wavenumber interval thus scales as $k^{-2/3}$, or $l^{2/3}$ per log-length interval in terms of spatial scales $l \sim k^{-1}$. We assume $\delta_B \lesssim 1$, because in practice, once $\delta_B \gg 1$, the random field effectively set a large-scale mean field of strength $B_0\sim \delta B$ in each coherence cell, so that such a situation would be similar to one where $\delta_B \sim 1$.

For a standard cascade, the energy transfer rate at mode $k$ can be written\footnote{We use the denomination $\gamma_{\ldots}$ to write turbulent transfer rates, injection into the cascade or damping on kinetic scales through wave-plasma interactions; the corresponding quantities are positive.} $\gamma_k = k\,\delta v_k$, where $\delta v_k \propto k^{-1/3}$ represents the characteristic eddy velocity. We assume equal magnetic and velocity fluctuation spectra, {\it viz.} $\delta v_k\propto {\mathcal E_B}_k^{1/2}$, which implies $\gamma_k \propto k {\mathcal E_B}_k^{1/2}$. In the inertial range, that is $k_{\rm inj}\ll k\ll k_{\rm kin}$ with $k_{\rm inj}$ the characteristic wavenumber at which energy is injected into the cascade and $k_{\rm kin}$ the wavenumber at kinetic scales where the cascade terminates by dissipation into plasma heating (not particle acceleration), the power spectrum ${\mathcal E_B}_k$ then emerges in the standard way as the steady-state solution to the cascade equation $\partial_t {\mathcal E_B}_k\,=\,-k\partial_k\left(\gamma_k {\mathcal E_B}_k\right)$. Up to factors of the order of unity, which we ignore here, $k_{\rm inj} \sim \ell_{\rm c}^{-1}$. To complete the cascade description, one needs to add an external reservoir of energy $\mathcal E_{\rm ext}$, which is injected at rate $\gamma_{\rm inj}$ into the cascade at the outer scale, thus contributing to turbulence driving, as well as a dissipation rate $\gamma_{\rm kin}$ at the dissipation scale.

To model the transfer of energy between turbulence and particles, we introduce a kernel $\Phi(k;\,p)$, which details how the energy gain $\Delta \mathcal E_p$ of the nonthermal population around momentum $p$ during time interval $\Delta t$ is distributed over wavenumbers $k$ of the cascade, meaning that a fraction $\Phi(k;\,p)$ of $\Delta \mathcal E_p$ is drawn from the part of the cascade around wavenumbers $k$. This function $\Phi(k;\,p)$ is normalized according to $\int {\rm d}\ln k\,\Phi(k;\,p)\equiv 1$. 

A common assumption is that particles of momentum $p$ and gyroradius $r_{\rm g} = pc/eB$ (with $B$ the total mean magnetic field strength) draw energy from turbulence at wavenumbers $k \sim r_{\rm g}^{-1}$, because particles of gyroradius $r_{\rm g}$ are insensitive to small-scale modes $l \sim k^{-1}\ll r_{\rm g}$, while large-scale modes $l\gg r_{\rm g}$ tend to renormalize the mean magnetic field along which particles move adiabatically. There are exception of course. In the context of quasilinear models, particle energization can occur both through gyroresonant interactions with Alfv\'en or magnetosonic modes ($k \sim r_{\rm g}^{-1}$), or transit-time damping interactions with compressive fluctuations, which are essentially non-resonant in wavenumber space ($k\lesssim r_{\rm g}^{-1}$)~\cite{1989ApJ...336..243S,98Schlick,2007MNRAS.378..245B,2008ApJ...684.1461Y,19Teraki,2020PhRvD.102b3003D}. Therefore, even in a quasilinear context, it proves difficult to write a unique wave-particle coupling kernel, unless one separates the cascade into its Alfv\'enic and compressive sub-cascades. To complicate somewhat the matter, the  inherent anisotropy of the cascade tends to suppress gyroresonant interactions with Alfv\'enic and slow magnetosonic modes, e.g.~\cite{2000PhRvL..85.4656C,2002PhRvL..89B1102Y}, unless additional effects such as resonance broadening are taken into account, see e.g.~\cite{14Lynn,18Xu} as well as~\cite{2020PhRvD.102b3003D} and references therein. Finally, in the generalized Fermi scenario of Ref.~\cite{2022PhRvL.129u5101L}, the particle gains energy by interacting with modes on scales $k\lesssim r_{\rm g}^{-1}$, although most of the energy gain appears to come from modes $k \sim r_{\rm g}^{-1}$.  

In order to sample those possibilities, while remaining as general as possible, we have considered two possible kernels, one describing gyroresonance, the other describing nonresonant interactions with larger-scale modes. Our gyroresonant kernel takes a Gaussian form 
\begin{equation}
\Phi(k;\,p)\propto\exp\left\{-\frac{1}{2}\left[\ln\left(k r_{\rm g}\right)\right]^2\right\}\,
\label{eq:kernel1}
\end{equation}
which formally describes a broadened resonance. As we have not observed a significant impact of the choice of kernel on the particle and wave spectra in our exploration, we restrict the presentation of our results to gyroresonant interactions and defer those corresponding to non-resonant interactions to Appendix~\ref{sec:appA}.

The turbulence cascade system, accounting for nonlinear feedback from particle acceleration can eventually be written in the form
\begin{align}
    \partial_t {\mathcal E_B}_k \,=\,& -k\partial_k\left(\gamma_k\,{\mathcal E_B}_k\right) - \int{\rm d}\ln p\, \Phi\left(k;\,p\right) \partial_t \mathcal E_p\nonumber\\
    & + \gamma_{\rm inj}\mathcal E_{\rm ext}k_{\rm inj}\delta\left(k-k_{\rm inj}\right)\nonumber\\
    & - \gamma_{\rm kin}\mathcal E_{k_{\rm kin}}k_{\rm kin}\delta\left(k-k_{\rm kin}\right)\,,
\label{eq:cascade}
\end{align}
The last two terms describe turbulence driving on the outer scale and dissipation on kinetic scales, as discussed before, while the first two describe advection in $k-$space by nonlinear mode-mode interactions within the turbulence and cascade damping by particle acceleration. For what concerns the inertial range, therefore, only the first two matter. The normalization $\int {\rm d}\ln k\,\Phi(k;\,p)\equiv 1$ guarantees that the total turbulent plus nonthermal particle energy is conserved up to the source due to external driving and the sink associated with dissipation into thermal plasma heating.  We have chosen here to describe the turbulent cascade using pure advection in wavenumber space, as in Ref.~\cite{2008JGRA..113.5103H}; other, more refined descriptions involving diffusion or diffusion-advection processes are possible, see e.g. the discussion in Ref.~\cite{1990JGR....9514881Z}. 

We now recall that the nonlinear interaction term $\gamma_k\propto k{\mathcal E_B}_k^{1/2}$. In practice, we thus define a reference rate $\gamma_0$ such that
\begin{equation}
    \gamma_k = \gamma_0\,a_\gamma\,k\,{\mathcal E_B}_k(t)^{1/2}\,, 
    \label{eq:gammadep}
\end{equation}
with normalization $a_\gamma\equiv 1/\left[k_{\rm inj}{\mathcal E_B}_{k_{\rm inj}}(t=0)^{1/2}\right]$ for dimensional reasons. 

Regarding the acceleration rate, we assume here that it scales in direct proportion to the amount of magnetic energy on the relevant scales, meaning
\begin{align}
    D_{pp}&\,=\,\nu_p\,p^2\,a_\nu\,\int {\rm d}\ln k\, \Phi(k;\,p)\,{\mathcal E_B}_k(t),\,\nonumber\\
    A_p &\,=\,\nu_p\,p\,a_\nu\,\int {\rm d}\ln k\, \Phi(k;\,p)\,{\mathcal E_B}_k(t)\,,
    \label{eq:nudep}
\end{align}
with normalization $a_\nu\equiv 1/\left[\int {\rm d}\ln k\, \Phi(k;\,p)\,{\mathcal E_B}_k(t=0)\right]$. There is no specific reason why the kernel $\Phi(k;\,p)$ that appears in Eq.~(\ref{eq:nudep}) should be the same as that which controls the feedback described by Eq.~(\ref{eq:cascade}). In the present framework, however, both should  retain the same essential general characteristics, therefore setting them to be equal appears as a reasonable choice. However, while the above linear dependency of $D_{pp}$ and $A_p$ on ${\mathcal E_B}_k$ holds both for the original Fermi scenario and for quasilinear models, a different relationship may arise if particles are accelerated in intermittent regions of strong velocity gradients~\cite{PhysRevD.104.063020,Bresci+22,2022PhRvL.129u5101L}. We will discuss possible consequences in Sec.~\ref{sec:caveat}. 

For reference, we mention that previous studies on this topic have adopted a Fokker-Planck description for the evolution of the particle distribution function, characterized by a diffusion coefficient $D_{pp}\propto p^q$ with $q\simeq 5/3$ extracted from quasilinear calculations in isotropic wave turbulence, used a wave damping term corresponding to exact gyroresonance, and described the cascade through diffusion in wavenumber space. In the present setting, the nonlinear features of stochastic acceleration are encoded in a general manner through the kernels describing the coupling between particles and the turbulence, and more specifically described by the dependencies of $\gamma_k$, $D_{pp}$ and $A_p$ on the time-dependent turbulent and particle energy contents.

\section{Implications for spectra}\label{sec:impl}
\subsection{A numerical example}
We integrate Eq.~(\ref{eq:cascade}) together with Eq.~(\ref{eq:model1}) describing the evolution of $\mathcal E_p$ in the Fokker-Planck model, or Eq.~(\ref{eq:model2}) for the power-law model, to obtain numerical estimates of the spectra accounting for the back-reaction that results from particle acceleration. For pedagogical purposes, we first do so with the same parameters as in Fig.~\ref{fig:nofdbck}. In detail, for a numerical time step $\delta t$, we choose here $\gamma_0\,\delta t = 2.6\times 10^4$ and $\nu_p\,\delta t=1.3$ for model 2, or $\nu_p\,\delta t=0.86$ for model 1. The dynamical range covers a large range of length scales extending from $0.3 r_{\rm g}(p_0)$ to $5\times 10^{10}r_{\rm g}(p_0)$, the largest scale setting the outer scale of turbulence $k_{\rm inj}^{-1}$. This choice of parameters is ad hoc, and intended to bring to light the main physical effects. Further below, we will update those parameters in order to make connection with realistic physical conditions. 

\begin{figure}[h]
\includegraphics[width=0.48\textwidth]{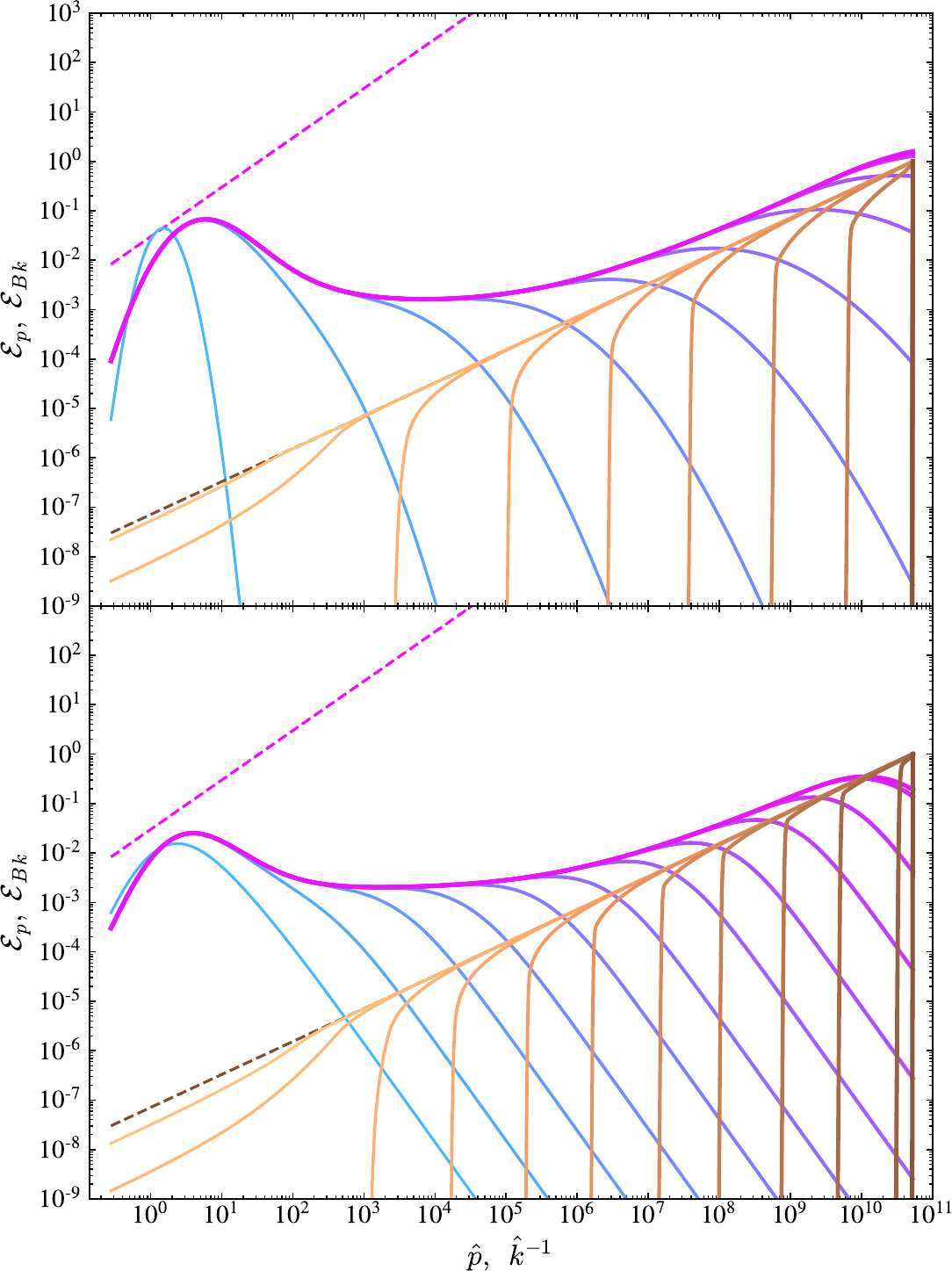}
\caption{Time-dependent solutions for the differential energy density spectrum $\mathcal E_p(p,t)\equiv 4\pi p^4 f(p,t)$ (blue colors), now accounting for turbulence damping by particle acceleration. Other parameters have remain unchanged with respect to those used in Fig.~\ref{fig:nofdbck}. The magnetic energy spectrum  ${\mathcal E_B}_k(t)$ is here shown at various times in orange colors. The abscissa is $\hat p\equiv p/p_0$ for $\mathcal E_p$, and $\hat k^{-1} = k^{-1}/r_{\rm g}(p_0)$ for ${\mathcal E_B}_k$.  Top panel: solution for the Fokker-Planck model described by Eq.~(\ref{eq:model1}); bottom panel: solution of the power-law model described by Eq.~(\ref{eq:model2}). Lines plotted in light-thin to deep-thick colors range from early to late times and are plotted at constant time intervals. The dashed purple line indicates a scaling $\mathcal E_p\propto p$, while the dashed orange line indicates the scaling expected for the turbulence cascade in the absence of nonlinear feedback, ${\mathcal E_B}_k\propto k^{-2/3}$. 
\label{fig:fdbck}}
\end{figure}

The results are shown in Fig.~\ref{fig:fdbck} and they are to be read as follows.
The energy spectra are plotted using the same conventions as in Fig.~\ref{fig:nofdbck}.
To ease the reading of the figure, we have chosen to plot the magnetic energy spectra alongside, as a function of spatial scale $\hat k^{-1}=k^{-1}/r_{\rm g}(p_0)$. The underlying motivation is to overlay both spectra and to better display the duality between $k$ and $p$ that emerges from the gyroresonance condition $\hat k^{-1}=\hat p$. The dashed orange line indicates the magnetic energy spectrum (per wavenumber log-interval) that would be observed in the absence of damping, ${\mathcal E_B}_k \propto k^{-2/3}$. 

Qualitatively, the general time evolution proceeds as follows. Shortly after injection, the accelerated particle spectrum can be read off as the light blue line to the bottom left of the figure, peaking at momenta $\hat p\sim 1$. Its evolution has drawn energy from the cascade at gyroresonant wavenumbers $\hat k^{-1}\sim 1$, thus eroding the power spectrum at those momenta, see the light orange line to the bottom left. This, in turn, slows down then eventually suppresses further acceleration of particles with momenta $\hat p\sim 1$. For this reason, further time-dependent solutions do not show substantial evolution at low momenta. Nevertheless, at any given time, particles of sufficiently large momenta, meaning in a part of the inertial range where the magnetic power spectrum has not suffered significant damping, can feed off the existing magnetic energy and undergo further acceleration. This contributes to damp the magnetic energy spectrum, as can be read off Fig.~\ref{fig:fdbck} by following the orange lines in time, from light-thin to deep-heavy. A direct comparison of both figures, with damping (Fig.~\ref{fig:nofdbck}) and without (Fig.~\ref{fig:fdbck}) demonstrates that acceleration feedback affects both energy spectra rather dramatically.

\subsection{Spectral shapes in the presence of damping}
We now discuss those spectral shapes in more quantitative details. Inside the inertial range, only the first two terms on the rhs of Eq.~(\ref{eq:cascade}) determine the evolution of ${\mathcal E_B}_k$: nonlinear mode coupling, which advects turbulent energy from large to small scales, and acceleration damping, which erodes the turbulence spectrum, acting in the opposite direction from small to large scales. At any given time $t$, damping is maximal at the wavenumber $k_{\rm d}(t)$ where those two rates meet, {\it i.e.} 
\begin{equation}
    \nu_p\mathcal E_{\rm p_{\rm d}} \sim \gamma_{k_{\rm d}} {\mathcal E_B}_{k_{\rm d}}\quad\quad\left({\rm damping}\right)\,.
    \label{eq:damping}
\end{equation}
This equation tacitly uses the correspondence $\hat p \leftrightarrow \hat k^{-1}$ to define $p_{\rm d}$ through $\hat p_{\rm d} = \hat k_{\rm d}^{-1}$, equivalently $r_{\rm g}(p_{\rm d}) = k_{\rm d}^{-1}$. If $\nu_p\mathcal E_{\rm p}\ll \gamma_{\rm k}{\mathcal E_B}_k$ at all $p$, assuming $\hat k \sim \hat p^{-1}$, then damping is ineffective and $\mathcal E_p$ increases as in the absence of feedback. Since $\mathcal E_p$ increases with time in that regime, while $\gamma_{\rm k}{\mathcal E_B}_k$ remains unchanged, the particle energy density may eventually increase up to the point where Eq.~(\ref{eq:damping}) becomes satisfied, at some time $t_{\rm d}$, and at a momentum $p_{\rm d}(t_{\rm d})$ that we write $p_{\rm pk}$. The subscript $_{\rm pk}$ underlines the fact that the peak of the particle energy density distribution will be found around $p_{\rm pk}$ at later times.

At later times, the damping rate is maximal at $k_{\rm d}$, because power has already been removed on smaller spatial scales $\hat k^{-1}\ll \hat k_{\rm d}^{-1}$ (equivalently $\hat p\ll \hat p_{\rm d}$), while on large spatial scales $\hat k^{-1}\gg \hat k_{\rm d}^{-1}$ (equivalently $\hat p\gg \hat p_{\rm d}$), the energy flows faster through the turbulence cascade than what can be removed by particles, so that ${\mathcal E_B}_k$ remains effectively unscathed.  In the numerical example presented in Fig.~\ref{fig:fdbck}, the numerical parameters at the initial time are such that the condition expressed by Eq.~(\ref{eq:damping}) above is met at $\hat p \simeq 5$, which indeed corresponds, to within a numerical prefactor of order unity, to where the particle energy spectrum stops evolving.

\begin{figure}[h]
\includegraphics[width=0.48\textwidth]{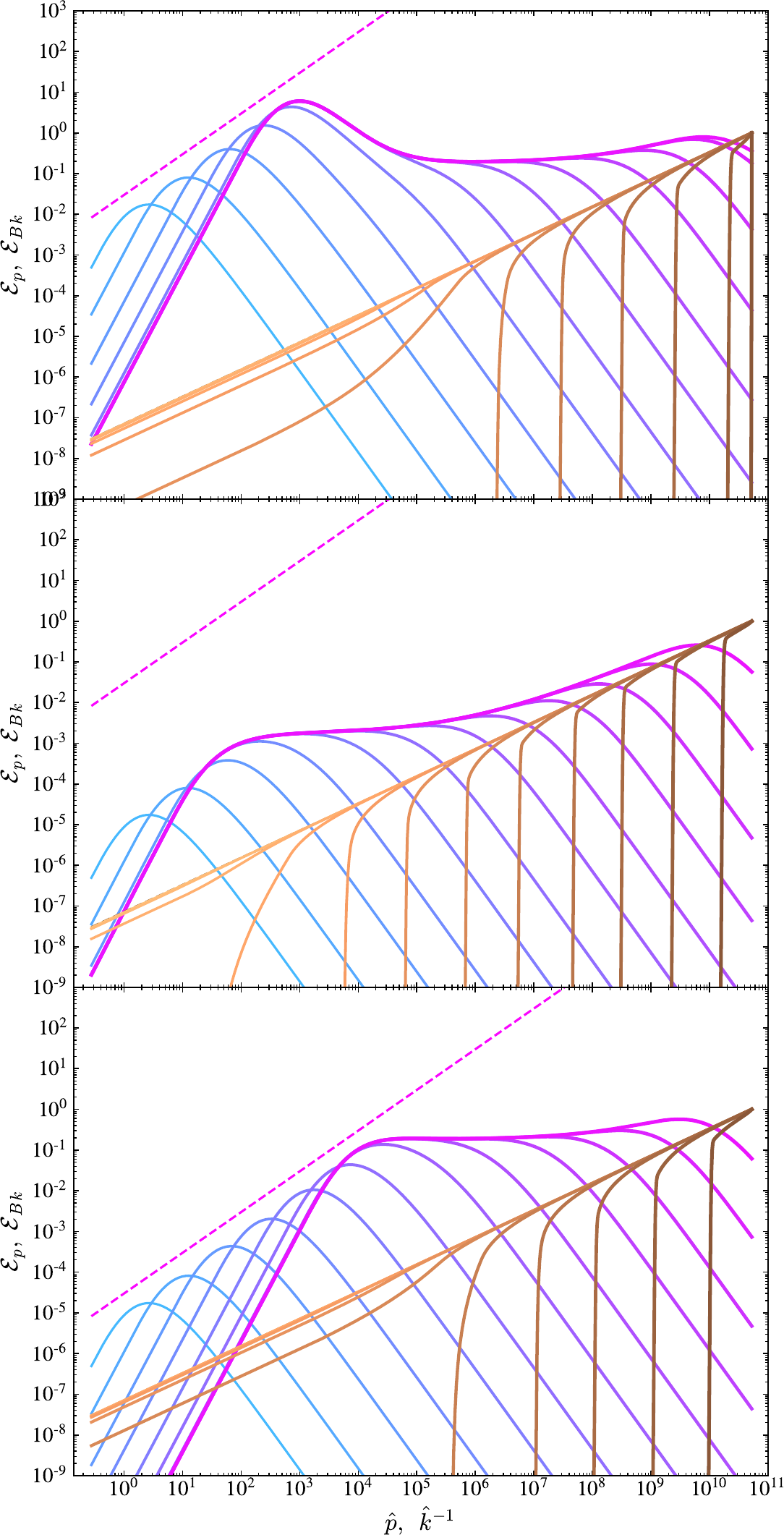}
\caption{Same as Fig.~\ref{fig:fdbck}, for different initial conditions. Top panel: cascade rate $\gamma_k$ increased by 100; middle panel: $\mathcal E_p(t=0)$ decreased by $10^3$; bottom panel: both $\gamma_k$ increased by 100 and $\mathcal E_p(t=0)$ decreased by $10^3$. Other parameters unchanged. Both panels consider the power-law model, formalized by Eq.~(\ref{eq:model2}). See text for details.
\label{fig:var}}
\end{figure}

Once damping becomes effective, {\it i.e.} once Eq.~(\ref{eq:damping}) is verified in some range of momenta, the general behavior of $\mathcal E_p(t)$ derives from two principles: momenta $p\gtrsim p_{\rm d}$ evolve at a rate fixed by the acceleration physics as in the absence of feedback, while the energy density of particles increases by as much as the turbulence can feed it at $k_{\rm d}$. In the above example, damping sets in soon after injection and the nonthermal particle energy density at $\hat p_{\rm d}$ ($\mathcal E_{p_{\rm d}}\sim 10^{-2}\rightarrow 10^{-1}$ in the units of Fig.~\ref{fig:fdbck}) far exceeds the turbulent energy density at $k_{\rm d}$ (${\mathcal E_B}_{k_{\rm d}}\simeq 10^{-7}$ in those same units). During a time interval $\Delta t$, therefore, the critical momentum $p_{\rm d}$ increases approximately as $\Delta \ln p_{\rm d}\simeq \nu_p\Delta t$, while the particle energy density changes only by $\Delta \mathcal E_{p_{\rm d}} = \mathcal E_{p_{\rm d}+\Delta p_{\rm d}} - \mathcal E_{p_{\rm d}} \simeq {\mathcal E_B}_{k_{\rm d}}$ (since $\Delta\ln k_{\rm d} \sim -\Delta\ln p_{\rm d}$), which is much smaller than $\mathcal E_{p_{\rm d}}$. Correspondingly, this process shapes a spectrum with an approximate scaling $\mathcal E_p(t)\propto p^0$ at $p_{\rm pk}<p<p_{\rm d}(t)$, which maintains the overall particle energy density roughly constant. In terms of spectral index, this corresponds to $s\sim 2$. This segment can be found in Fig.~\ref{fig:fdbck} in the range $\hat p \sim 10^2$ to $\hat p \sim 10^6$ roughly.

As discussed thereafter (Sec.~\ref{sec:physpar}), $\mathcal E_{p_{\rm d}}$ at $t_{\rm d}$ is expected to be significantly larger than ${\mathcal E_B}_{k_{\rm d}}$. Consequently, as $\mathcal E_p$ develops a flat spectrum at times $t>t_{\rm d}$ and momenta $p>p_{\rm pk}$, while ${\mathcal E_B}_k$ grows as $\hat k^{-2/3}$, $\mathcal E_{p_{\rm d}(t)}$ may eventually become comparable to that in the turbulent cascade at $k_{\rm d}(t)$. Henceforth, we refer to the momentum at which this occurs as $p_\times$. In the range $p> p_\times$, the spectral shape transits to $\mathcal E_p\propto p^{0.5-0.7}$ (corresponding to an index $s\simeq 1.3-1.5$ for ${\rm d}n/{\rm d}p\propto p^{-s}$) because the amount of turbulent energy density damped in the interval $\Delta\ln k \simeq -\Delta\ln p$ in a time interval $\Delta t$ is now given mostly to the particles, instead of cascading to smaller length scales. Correspondingly, the particle then develops a spectrum enslaved\footnote{Above, we wrote $\mathcal E_p\propto p^{0.5-0.7}$, not $\mathcal E_p\propto p^{2/3}$, both to reflect the fact that the index is not exactly $q_B-1$ with $q_B=5/3$ the index of the turbulent power spectrum, and because $q_B$ itself can take values between $3/2$ and $5/3$.} to that of ${\mathcal E_B}_k$, as is clearly apparent in Fig.~\ref{fig:fdbck}.

The cases shown in Fig.~\ref{fig:fdbck} make ad hoc choices of the overall normalization of the two parameters that determine advection, $\nu_p$ and $\gamma_k$, as of the initial energy density in particles and magnetic field. The general evolution depicted above nevertheless offers a generic overview of the phenomenology. We illustrate this point in Fig.~\ref{fig:var} by integrating the system for the power-law model with different parameters: in the top panel, the cascade rate $\gamma_k$ has been multiplied by $100$, while in the middle panel, the initial nonthermal particle energy density has been divided by $1000$, and in the bottom panel, both the cascade rate has been increased by $100$ and the initial particle energy density has been divided by $1000$; all other parameters have remained unchanged. Increasing $\gamma_k$ or decreasing $\mathcal E_p(t=0)$ delays the onset of damping, according to Eq.~(\ref{eq:damping}). Consequently, damping is weak at initial times in both top and middle panels, and all the more so in the lower panel. Particle acceleration then proceeds, at least initially, as in Fig.~\ref{fig:nofdbck}, meaning that the momentum at which the particle energy distribution peaks shifts as $\exp\left(\nu_p t\right)$, while the energy density increases in direct proportion to that peak momentum. Once damping sets in, this peak momentum stops evolving and the segments discussed above emerge.

The general shape of the time-dependent spectrum can thus be summarized in the following way. At early times $t< t_{\rm d}$, damping has not set in and particles are accelerated as described in Sec.~\ref{sec:linear}. In particular, the spectrum can be obtained
as the solution to the transport equation, {\it e.g.} Eq.~(\ref{eq:model1}) or (\ref{eq:model2}). At later times, $t>t_{\rm d}$, when damping sets in, the spectral shape is characterized by three characteristic momenta: $p_{\rm pk}$, constant in time and defined as  the momentum at which $\mathcal E_p$ peaks at time $t_{\rm d}$, {\it i.e.} $p_{\rm pk}\simeq p_{\rm d}(t_{\rm d})$; $p_{\rm d}(t)$, the momentum that sets the scale $\hat k_{\rm d}=\hat p_{\rm d}^{-1}$ up to which damping is effective; finally, $p_\times$, the momentum at which $\mathcal E_{p_\times}$ becomes comparable to the turbulent magnetic energy density on scale $\hat k_\times = \hat p_\times^{-1}$. To a reasonable approximation, $p_{\rm d}(t)$ can be defined as the characteristic momentum at which $\mathcal E_p$ would peak in the absence of damping, because damping does not affect the acceleration rate of particles of momentum $p_{\rm d}$. The spectral shape at times $t>t_{\rm d}$ is thus characterized by the following segments. First consider the case $p_{\rm d}(t) < p_\times$. For $p< p_{\rm d}(t)$, the spectrum is frozen in time, and characterized by a break at momentum $p_{\rm pk}$: below $p_{\rm pk}$, the spectral shape corresponds to the low-energy part of the spectrum obtained by linear evolution (meaning, in the absence of damping) at time $t_{\rm d}$; in the range  $p_{\rm pk}<p<p_{\rm d}(t)$, the spectrum is flat, characterized by a power-law of index $s\simeq 2$. For $p> p_{\rm d}(t)$, the spectral shape is that obtained by linear evolution up to time $t$, because particles in that range of momenta have interacted with undamped modes of the cascade at all times. If $p_{\rm d}(t)>p_\times$, an extra segment appears. It is characterized by a new power-law spectrum with index $s\simeq 1.3-1.5$ in the range $p_\times < p < p_{\rm d}(t)$; the flat spectrum with index $s\simeq 2$ extends from $p_{\rm pk}$ to $p_\times$; other segments remain unchanged.

Depending on the value of $p_0$ and the initial $\mathcal E_p$, the duration -- equivalently, extent in dynamic range -- of the linear phase can be more or less pronounced, as illustrated by Figs.~\ref{fig:fdbck}  and \ref{fig:var}. We stress that the above spectral shapes are time-dependent and that they assume an impulsive injection of particles at momentum $p_0$ at the initial time. The influence of the driving timescale and the possible continuous injection of particles on the overall time-integrated spectra will be discussed in Sec.~\ref{sec:tstir} and \ref{sec:loss}.

\subsection{Connection to physical parameters}\label{sec:physpar}
To make contact with the quantities that characterize the turbulence, we first recall that $\gamma_{k_{\rm inj}}\sim v_{\rm A}/\ell_{\rm c}$ and ${\mathcal E_B}_{k_{\rm inj}}\sim \mathcal E_B \sim \delta_B^2 \mathcal E_{B_0}\sim \delta_B^2 v_{\rm A}^2 \rho$, where $\rho$ represents the mass density of the thermal plasma. In the following, we use $\beta_{\rm A}\equiv v_{\rm A}/c$. Alternatively, one could use $\mathcal E_{B_0}\sim \beta^{-1} n T$,  to relate the magnetic energy density to the plasma beta parameter, its density $n$ and temperature $T$. 

Unfortunately, the scaling of the advection rate $\nu_p$ is not as well estimated. In quasilinear theory, one expects $\nu_p \simeq \beta_{\rm A}^q \delta_B^2 c/\ell_{\rm c}$ with $q=2$ or $q=3$ depending on the acceleration mechanism and the source of resonance broadening, as discussed in Ref.~\cite{2020PhRvD.102b3003D}, while in the generalized Fermi model of Ref.~\cite{2022PhRvL.129u5101L}, one expects $q=3$ in the sub-relativistic regime. Here, we retain a general scaling characterized by $q$.

To simplify the following estimates, we assume that, at time $t_{\rm d}$ at which damping first sets in, most of the nonthermal energy density is concentrated at some momentum $p_{\rm pk}$ and neglect the details of the spectral shape of the spectrum beyond $p_{\rm pk}$. This is a reasonable assumption because $\gamma_k {\mathcal E_B}_k$ does not depend on $k$ (per Kolmogorov universality assumption), so that Eq.~(\ref{eq:damping}) is first verified at the point where $\mathcal E_p$ peaks. Furthermore, we write the total nonthermal particle energy density as $\mathcal E_{\rm nth}(t)$ and thus assume that $\mathcal E_{\rm nth}\sim \mathcal E_{p_{\rm pk}(t)}$ at time $t<t_{\rm d}$, extending the definition of $p_{\rm pk}$ to $p_{\rm pk}(t)$ at $t<t_{\rm d}$, corresponding to the momentum at which $\mathcal E_p$ peaks in the absence of damping. Recalling first that $\mathcal E_{p_{\rm pk}(t)}(t)\simeq \mathcal E_{\rm nth,0}\,p_{\rm pk}(t)/p_0$ in that regime with $\mathcal E_{\rm nth,0}\equiv \mathcal E_{\rm nth}(t=0)$, recalling also that $\gamma_k {\mathcal E_B}_k = \gamma_{k_{\rm inj}} {\mathcal E_B}_{k_{\rm inj}}$, the condition Eq.~(\ref{eq:damping}) that determines the onset of damping can be written as
\begin{align}
      \mathcal E_{\rm nth}&\,\sim\, \beta_{\rm A}^{3-q} \rho c^2\,\sim\, \beta_{\rm A}^{1-q}\beta^{-1} nT\, \sim\, \beta_{\rm A}^{1-q} \mathcal E_{B_0}\,.
    \label{eq:physdamp}
\end{align} 
This critical value is written $\mathcal E_{\rm nth,\,nlin}$ in the following.
This is neither a trivial nor an intuitive result, in the sense that $\mathcal E_{{\rm nth}}$ can exceed $\mathcal E_{B_0}$ significantly before feedback occurs, all the more so if $\beta_{\rm A}\ll1$. The reason is  that, as particles get accelerated, energy keeps being injected in the turbulence cascade on the outer scale, so that what truly governs the amount of energy that accelerated particles can receive is the energy flux along the cascade as integrated over time, not the initial energy content on the outer scale. This, of course, requires turbulence to be driven over sufficiently long times. Assuming that acceleration proceeds exponentially fast at rate $\sim\nu_p$, this corresponds to a driving time
\begin{equation}
    t_{\rm inj}\gtrsim \frac{1}{\nu_p}\ln\left(\mathcal E_{\rm nth,\,nlin}/\mathcal E_{\rm nth,0}\right)\,.
    \label{eq:drivetime}
\end{equation}

Alternatively, one can write the momentum $p_{\rm pk}$ at which damping becomes important as
\begin{align}
     \frac{p_{\rm pk}}{p_0}&\,\sim\, \beta_{\rm A}^{3-q} \frac{\rho c^2}{\mathcal E_{\rm nth,0}}\,\sim\, \beta_{\rm A}^{1-q}\beta^{-1} \frac{nT}{\mathcal E_{\rm nth,0}}\nonumber\\
     &\, \sim\, \beta_{\rm A}^{1-q} \frac{\mathcal E_{B_0}}{\mathcal E_{\rm nth,0}}\,.
    \label{eq:physdamp_p}
\end{align}

In the numerical example presented earlier in Figs.~\ref{fig:fdbck} and \ref{fig:var}, the ad hoc values of $\nu_p$ and $\gamma_0$ were not fixed to $v_{\rm A}$ and other quantities, which explains why feedback occurred at values of $\mathcal E_p$ sometimes below that of $\mathcal E_B$, except for the top or bottom panels of Fig.~\ref{fig:var}, which present a case that satisfies Eq.~(\ref{eq:physdamp}) at $\delta_B\sim1$ and $\beta_{\rm A}\sim 1$. This choice was dictated by numerical constraints guaranteeing the accuracy and stability of the integration. Nevertheless, the observed generic behavior and its interpretation given above allows to extrapolate those results to realistic cases of interest.

If $\mathcal E_{\rm nth,0} \ll \mathcal E_{\rm nth,\,nlin}$ initially, acceleration proceeds as in the absence of damping until damping sets in, provided the dynamic range of the cascade as well as all other external constraints on the system (lifetime, age etc.) allow it. At this point, the energy density is assumed to peak at $p_{\rm pk}$ as determined by Eq.~(\ref{eq:physdamp_p}); by definition, $p_{\rm d}\sim p_{\rm k}$ at that time. Furthermore, Eq.~(\ref{eq:physdamp}) guarantees that 
$\mathcal E_{p_{\rm k}}\gg {\mathcal E_B}_{k_{\rm d}}$, hence the spectrum develops a spectrum ${\rm d}n/{\rm d}p\propto p^{-2}$  up to the maximum energies, starting at momentum $p_{\rm pk}$. While $p_{\rm d}(t)$ keeps evolving in time afterwards, $p_{\rm pk}$ remains approximately unchanged because the turbulence on scales $\gtrsim r_{\rm g}(p_{\rm pk})$ is being  damped. The regime $\mathcal E_p \propto p^{0.5-0.7}$ may eventually emerge, although only in a restricted range of gyroradii not far below the outer scale, because $\mathcal E_{p_{\rm d}}\gtrsim \mathcal E_B$ for nearly all $p_{\rm d}$, according to Eq.~(\ref{eq:physdamp}). 

It may be worthwhile to point out that Eq.~(\ref{eq:physdamp}) also indicates that the nonthermal particle energy density should remain at all times less or at most comparable to $\rho c^2$, which implies that the Alfv\'en velocity should not be modified significantly by particle acceleration. This justifies our initial hypothesis of constant $\beta_{\rm A}$. 
To generalize those relations to plasmas of relativistic temperature, one needs to operate the substitution $\rho c^2 \rightarrow e$ (up to order unity prefactors), writing $e$ as the total plasma energy density, and $v_{\rm A}\rightarrow u_{\rm A}\sim \left(\mathcal E_{B_0}/e\right)^{1/2}$, with $u_{\rm A}$ the Alfv\'en $4-$velocity, meaning that $\beta_{\rm A}$ interpreted as $u_{\rm A}/c$ can now take values larger than unity.

\section{Discussion}\label{sec:disc}

\subsection{Limitations and caveats}\label{sec:caveat}
The present discussion remains exploratory in nature, because of the numerous uncertainties that plague the physics of particle acceleration in magnetized turbulence, even at the test-particle level. Nonetheless, we first wish to stress that the assumption of gyroresonance that underlies the definition of the kernel $\Phi(k;\,p)$ plays only a minor role in our finding. 
The key point that controls the evolution of the spectra in Fig.~\ref{fig:fdbck} is that turbulent energy flows through the cascade from right to left (large to small scales), while the erosion of that cascade progresses from left to right. The reason why this occurs, in turn, is that the time-dependent spectrum of accelerated particles in the absence of feedback has a high-energy slope $s > 2$, 
which implies that, as advection shifts that spectrum to the right, erosion also shifts to the right. This explains, in particular, why the present results remain mostly insensitive to the shape of the kernel. Of course, one could conceive a situation in which particles are accelerated by drawing energy from the largest scales of the cascade only. This however requires a rather fine-tuned kernel $\Phi(k;\,p)$, strongly peaked in $k$ at $k_{\rm inj}$;  otherwise the previous results are recovered, as we have checked. 

On the other hand, the relationship [Eq.~(\ref{eq:nudep})] that relates $D_{pp}$ and $A_p$ to the magnetic energy content does play a non-trivial role in regulating the damping of the cascade. So far, we have assumed that the acceleration rate depends linearly on ${\mathcal E_B}_k$, as motivated by the original Fermi scenario or quasilinear models. This implies that acceleration proceeds as long as there remains magnetic energy, which implies, by virtue of Eq.~(\ref{eq:cascade}) that damping persists until complete erosion of the turbulent power spectrum on the relevant scales. Now, if particle acceleration rather proceeds through the interaction of particles with localized, intermittent structures~\cite{PhysRevD.104.063020,Bresci+22,2022PhRvL.129u5101L}, one should anticipate that particle acceleration would remove power from those localized structures only, not from the average turbulent bath where acceleration is slow or ineffective. Acceleration would then stall, hence damping would cease, even though the magnetic energy on those scales has not been fully removed. 

\begin{figure}[h]
\includegraphics[width=0.48\textwidth]{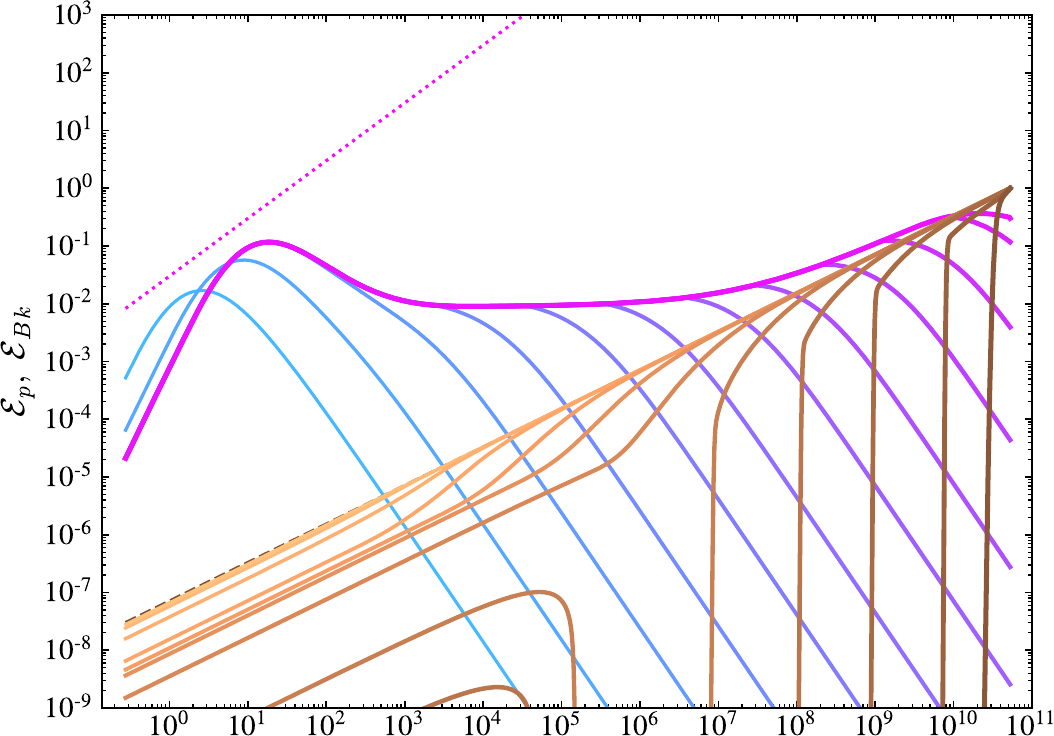}
\caption{Same as Fig.~\ref{fig:var} (top panel), assuming however that the acceleration rate scales as a higher power of ${\mathcal E_B}_k$, namely $A_p = \nu_p\, \varphi_B^n$ with $n=3$ as described in the text. The case shown in Fig.~\ref{fig:fdbck} corresponds to $n=1$. The main consequence is to moderate the damping of the turbulence cascade, while the shape of the particle energy spectrum remains preserved. See text for details.
\label{fig:damping-int}}
\end{figure}

As a naive representation of this model, we replace the dependency $A_p = \nu_p\, \varphi_B$, where $\varphi_B = a_\nu \int {\rm d}\ln k\,\Phi(k;\,p){\mathcal E_B}_k$ by $A_p = \nu_p\, \varphi_B^n$ with $n>1$. This model amplifies the impact of damping on acceleration rate, in such a way as to quench acceleration rapidly once damping sets in, {\it i.e.} when $\varphi_B <1$. As illustrated by Fig.~\ref{fig:damping-int}, damping indeed becomes weak, at least at early times. The particle energy distribution however retains the same features as in the models discussed so far ($n=1$), because the acceleration rate has dropped sharply in the regions where damping has occurred, while particle acceleration still proceeds in undamped regions of the cascade. We have verified that different parametrizations, {\it e.g.} $A_p = \nu_p\, {\rm max}\left\{0,2\varphi_B-1\right\}$, which halts acceleration once the turbulent energy content on the relevant scales has dropped by half, would lead to a similar phenomenology.


\subsection{Accompanying acceleration processes}\label{sec:secondacc}
One must also question whether the bulk of accelerated particles cannot exert a positive feedback on the turbulence, instead of just acting as a sink for turbulent energy. 
Natural possibilities that spring to mind are instabilities generated by anisotropies in the particle momentum distribution. For instance, large-scale compression and shear in high-beta plasmas generate anisotropies that can excite firehose and mirror instabilities, which in turn build magnetic power on short (kinetic) length scales, thereby bypassing the usual scale-to-scale cascade~\cite{2023PhRvX..13b1014A}. 
This phenomenology pertains to high-beta plasmas, because the level of compression needed to reach the firehose or mirror thresholds, $\left\vert\Delta P/P\right\vert\gtrsim \mathcal O\left(1/\beta\right)$ with $P=nT$ the pressure and $\Delta P$ the amount of pressure anisotropy, remains moderate in that regime. Although we have assumed $\beta \lesssim 1$, the beta parameter $\beta_{\rm nth}$ of the nonthermal population must have reached substantial values if stochastic acceleration has entered a nonlinear regime: $\beta_{\rm nth}\sim \beta_{\rm A}^{1-q}$ according to Eq.~(\ref{eq:physdamp}). Hence, in principle, those nonthermal particles could induce firehose or mirror instabilities that would redistribute some energy on small length scales. 
To our knowledge, the development of such instabilities in a two-component plasma (thermal + nonthermal) has not received attention so far. 
Speculating on the consequences, one might expect that the added small-scale power could contribute to particle scattering, although not as effectively as the original cascade on scales $\gtrsim r_{\rm g}(p_{\rm d})$. Nevertheless, this would lead to acceleration, for instance through a form of magnetic pumping, combining scattering on small-scale inhomogeneities with adiabatic compression in the large-scale perturbations~\cite{Lichko_2017,2023ApJ...947...89L}. 

Magnetic reconnection in the damped part of the turbulence spectrum might also open another acceleration channel. Particle-in-cell simulations have demonstrated that in a (relativistic) magnetized turbulent plasma, reconnection acts as an injector of thermal particles into the Fermi acceleration process~\cite{17Zhdankin,2018ApJ...867L..18Z,18Comisso,2019ApJ...886..122C,2020ApJ...893L...7W,2021ApJ...922..172Z,2021ApJ...921...87N}. 
In detail, particles swept by current sheets are pre-accelerated up to a momentum such that their gyroradius exceeds the current sheet thickness, at which point they start exploring the larger scales of the turbulence cascade. In the present context, however, one may wonder whether turbulence damping below a macroscopic scale $k_{\rm d}^{-1}$ could not generate current sheets on that scale that would energize particles up to momentum $\sim p_{\rm d}$. Such a scenario should be investigated with dedicated numerical simulations.

Overall, it seems natural to expect that some form of secondary, residual acceleration could contribute to particle energization in the region of scales where turbulence has been damped. Provided this form of acceleration is less efficient than the primary channel of stochastic acceleration discussed earlier, the overall spectral shape with $s\sim 2$ would remain unchanged. As we have checked, the dominant effect is to shape a hard spectrum at momenta $p\lesssim p_{\rm pk}$ while shifting $p_{\rm pk}$ to larger values, slowly in time.

\subsection{Influence of the driving timescale}\label{sec:tstir}
So far, we have assumed that the timescale $t_{\rm inj}$ over which the turbulence is stirred far exceeds the other relevant timescales, $t_{\rm acc}\sim \nu_p^{-1}$ and $t_{\rm dec}\sim \gamma_{k_{\rm inj}}^{-1} \sim \ell_{\rm c}/v_{\rm A}$ in particular. In the opposite limit $t_{\rm inj}\lesssim t_{\rm dec}$, the turbulence can be regarded as decaying. Particle acceleration becomes ineffective if $t_{\rm dec}\ll t_{\rm acc}$, because most of the turbulent energy is converted to plasma heating before acceleration can occur. Such a situation occurs rather generically if the turbulence is of small amplitude ($\delta_B \ll1$) or sub-relativistic ($v_{\rm A}\ll c$), because the acceleration timescale increases fast with decreasing $\delta_B$ and $v_{\rm A}$. Oppositely, particle acceleration can proceed for a few turn-over times $t_{\rm dec}$ and generate a power-law spectrum with characteristic index $s \simeq 2\ldots 4$ in the large-amplitude, semi-relativistic regime of decaying turbulence~\cite{18Comisso,2019ApJ...886..122C}. Once the turbulence has decayed, acceleration stops and the spectra freeeze. Similar spectral indices are observed if the turbulence is stirred over $\sim 10 \ell_{\rm c}/v_{\rm A}$~\cite{17Zhdankin,2018ApJ...867L..18Z,2020ApJ...893L...7W,2021ApJ...922..172Z,2021ApJ...921...87N}. 

It appears natural to ask if the accelerated particles are able to back-react on the turbulence on such short timescales. At sub-relativistic Alfv\'enic velocities, the response is likely no, according to Eq.~(\ref{eq:physdamp}). On the numerical side, test-particle tracking in time-dependent MHD simulations has shown that in this limit ($v_{\rm A}\simeq 0.4\,c$, $\delta_B\gg 1$), the nonthermal spectrum is characterized by an index $s\simeq 4$~\cite{2022PhRvL.129u5101L}, in good agreement with that measured in PIC numerical simulations in relatively similar conditions ($v_{\rm A}\simeq 0.1\,c$, $\delta_B \gtrsim 1$)~\cite{2022ApJ...936L..27C}. This corroborates the idea that back-reaction through particle acceleration is weak in that regime, at least on the timescales probed, as otherwise, the spectra observed by following test-particles (in the MHD simulation) would have differed from those measured in PIC simulations, which by construction account for all nonlinear back-reaction effects. 

In the relativistic limit ($v_{\rm A}\simeq c$), Eq.~(\ref{eq:physdamp}) suggests that backreaction could be effective even on a few eddy turn-over timescales. In PIC numerical simulations, the ``nonthermal particle population'' cannot be distinguished from the thermal plasma, because the accelerated particles are extracted out of that pool and the energy densities in both sub-populations are comparable.  Interestingly, these simulations have reported spectra with index $s\simeq 2$ at $\delta_B\gtrsim 1$. This suggests that maybe, the true difference between the spectral indices observed in these simulations in the relativistic {\it vs} mildly or sub-relativistic limit stems from nonlinear feedback, not from a fundamental difference in the acceleration physics. A comparison of test-particle tracking in MHD simulations of relativistic turbulence with the above PIC simulations would provide an unambiguous test of that possibility. Another telltale signature would be to measure an acceleration timescale that increases with time at a fixed momentum, as well as an eroded power spectrum on short length scales. The degree of erosion of the power spectrum is however subject to the uncertainties affecting the relationship between acceleration rate and magnetic energy content, as discussed in Sec.~\ref{sec:caveat}. Turning the argument around, we remark that a detailed study of the time evolution of those various quantities (acceleration timescale, power spectrum etc.) could bring valuable information on the physical process governing particle acceleration in magnetized turbulence.

Damping of the power spectrum seems to have been observed in recent PIC simulations of relativistic turbulence~\cite{2023arXiv230111327G}, although in a different regime from that studied here. That work indeed focuses on radiative turbulence, in a regime such that particles cool through inverse Compton interactions on a short timescale, significantly shorter than the eddy turn-over time. Consequently, particles are not accelerated to large energies and damping is restricted to the smallest length scales of the cascade. 

Finally, we would like to draw attention to recent efforts that aim at deriving the shape of the accelerated particle spectrum from arguments of (generalized) maximum entropy in a collisionless turbulent plasma~\cite{2022PhRvX..12c1011Z,2022JPlPh..88c1703Z,
2022JPlPh..88e9201E,2023arXiv230403715E}. By construction, such studies focus on the long timescales, sufficiently long at least for relaxation to take place in the relevant region of the energy distribution. Quite interestingly, spectra with indices $s\sim 2$ then emerge as attractors. It is thus tempting to speculate that the nonlinear feedback process described here  represent a step on the route to these asymptotic spectra.

\subsection{Consequences of continuous injection and effects of losses and escape}\label{sec:loss}
For pratical purposes, one may also wonder how time-integrated or steady-state spectra look like when particles are continuously injected and how the present picture would be modified in the presence of losses through radiative interactions, escape in position space, or further injection of particles, which are common features of astrophysical phenomenological models.

So far, we have considered an impulsive injection of a certain spectrum (e.g., a broken power-law) at $t=0$, so that the calculated spectra effectively represent Green's functions in time. If particles are injected continuously, then one should integrate those Green's functions over the injection time history. In detail, if particles are injected at momentum $p_0$ at a rate $\nu_{\rm inj}$, then the time-integrated spectrum up to time $t$ is given by $\mathcal E^{\rm int}_p(t)\equiv \nu_{\rm inj}\int^t{\rm d}\tau\,\mathcal E_p(\tau)$. We provide here an expression for the general spectral shape, considering model 2 for simplicity (the results can be directly generalized to model 1). At times $t<t_{\rm d}$, nonlinear effects can be neglected, and
\begin{align}
\mathcal E^{\rm int}_p(t) \,\simeq\,& \frac{\nu_{\rm inj}}{\nu_p}\, \frac{p}{p_0}\, \mathcal E_{\rm nth,0}\quad \left[p< p_{\rm pk}(t)\right],\nonumber\\
\mathcal E^{\rm int}_p(t) \,\simeq\,& \frac{\nu_{\rm inj}}{\nu_p}\,\frac{p_{\rm pk}(t)}{p_0}\,\mathcal E_{\rm nth,0}\left[\frac{p}{p_{\rm pk}(t)}\right]^{-s_{\rm hi}+2}\quad \left[p> p_{\rm pk}(t)\right]\,,
\label{eq:Eplin}
\end{align}
with $p_{\rm pk}(t)$ the characteristic momentum at which the instantaneous spectrum $\mathcal E_p$ peaks. For model 2, $p_{\rm pk}(t)\simeq p_0\exp(\nu_p t)$, while for model 1, $p_{\rm pk}(t)\simeq p_0\exp(5\nu_p t)$ at $t<t_{\rm d}$. It is understood that, at times $t> t_{\rm d}$, $p_{\rm pk}(t)\simeq p_{\rm pk}$ as defined in Eq.~(\ref{eq:physdamp_p}). 

On longer timescales $t>t_{\rm d}$, back reaction acts in two ways: it freezes acceleration at $p<p_{\rm pk}$ and shapes a flat spectrum in the momentum range $p_{\rm pk}<p<p_{\rm d}(t)$, as discussed earlier. We recall that $p_{\rm d}(t)$ corresponds to the momentum at which $\mathcal E_p$ would peak in the absence of damping, {\it i.e.} $p_{\rm d}(t) \simeq p_{\rm pk}\exp(\nu_p(t-t_{\rm d}))\simeq p_0\exp(\nu_p t)$ (for model 2). Then
\begin{align}
    \mathcal E^{\rm int}_p(t) \,\simeq\,& \nu_{\rm inj}\mathcal E_{\rm nth,0}\left[\frac{1}{\nu_p}\frac{p}{p_0}  + \left(t-t_{\rm d}\right)\frac{p_{\rm pk}}{p_0}\,\left(\frac{p}{p_{\rm pk}}\right)^{-s_{\rm li}+2}\right]\nonumber\\ & + \nu_{\rm inj}\,\mathcal E_{\rm nth,0}\,\left(t-t_{\rm d}\right)\,p_0\,\delta(p-p_0)\quad \left[p< p_{\rm pk}\right],\label{eq:Epnonlin1}\\
\mathcal E^{\rm int}_p(t) \,\simeq\,& \nu_{\rm inj}\mathcal E_{\rm nth,0}\frac{p_{\rm pk}}{p_0}\Biggl\{
    \nonumber\\& \quad\Theta\left[p_{\rm d}(t)-p\right]\left[t-t_{\rm d}\,-\frac{\ln(p/p_{\rm pk})}{\nu_p} \right]
\nonumber\\ & + \frac{\Theta\left[p-p_{\rm d}(t)\right]}{\nu_p}\left(\frac{p}{p_{\rm d}(t)}\right)^{-s_{\rm hi}+2}\Biggr\}\quad \left[p> p_{\rm pk}\right]\,.
\label{eq:Epnonlin2}
\end{align}
As the general aim here is to capture the general shape of the time-integrated spectrum, numerical prefactors of the order of unity have been replaced by $1$ in these equations and only leading terms have been kept.

The second term in the bracket of Eq.~(\ref{eq:Epnonlin1}) captures the low-energy extension of the spectrum that freezes at late time due to turbulent damping. The corresponding power-law index has been set to $s_{\rm li}=-2$ by choice in the rest of the text (see Sec.~\ref{sec:linear}), but different models could lead to different values. In particular, the $p-$dependency would differ in model 1. Additionally, any secondary or residual acceleration would harden the spectrum in that range. A conservative interpretation of Eq.~(\ref{eq:Epnonlin1}), therefore, should limit itself to the observation that the time-integrated spectrum in the range $p<p_{\rm pk}$ must be harder than $s\simeq 1$ (equivalently $\mathcal E_p \propto p$). 

In Eq.~(\ref{eq:Epnonlin2}), the first term $\propto \Theta\left[p_{\rm d}(t)-p\right]$ (together with $p>p_{\rm pk}$), $\Theta(x)$ denoting the Heaviside distribution, concerns the region of momenta whose spectrum has been frozen due to turbulent damping. The first term $\propto t$ provides the dominant contribution at late times. The second term $\propto \Theta\left[p-p_{\rm d}(t)\right]$ instead represents the part of spectrum that gets accelerated in undamped regions of the turbulent cascade. Correspondingly, $\mathcal E^{\rm int}_p(t)$ retains the scaling $\propto p^{-s_{\rm hi}+2}$ of the high-energy tail of the injected spectrum. We have omitted here, for simplicity, the possible contribution that would arise from the segment $\mathcal E_p\propto p^{0.5-0.7}$, at momenta $p$ larger than $p_\times$, defined as that at which $\mathcal E_p$ becomes comparable to ${\mathcal E_B}_{k_\times}$ for $\hat k_\times = \hat p_{\times}^{-1}$. It would provide a contribution $\propto \nu_{\rm inj}\mathcal E_{\rm nth,0}(p_{\rm pk}/p_0)(t-t_\times)(p/p_\times)^{0.5-0.7}$ in the range $p_\times < p < p_{\rm d}(t)$, where $t_\times$ denotes the time at which $p_{\rm d}(t)=p_\times$.

In summary, therefore, the time-integrated spectrum preserves the general shape of the time-dependent spectra $\mathcal E_p$ illustrated in Figs.~\ref{fig:fdbck} and \ref{fig:var}, except for the low-energy segment below the peak momentum, which takes a power-law form with $s\simeq 1$ in the absence of damping ($t<t_{\rm d}$), and possibly harder at times $t> t_{\rm d}$.

The final term on the r.h.s. of Eq.~(\ref{eq:Epnonlin1}) accounts for particles that have been injected at $p_0$ at times $t>t_{\rm d}$, and that have not been accelerated because of the lack of turbulent power on the relevant scales. In some physical situations, this term might be absent altogether. Consider for instance a system in which particles are injected at one spatial boundary, then advected away at the same time as they are subject to stochastic acceleration in the comoving plasma turbulence. Integration over space of the particle energy density spectrum would then scale as the integral over time of the impulsive time-dependent spectra calculated earlier, so that one would recover Eq.~(\ref{eq:Epnonlin1}) up to that contribution $\propto \delta(p-p_0)$.

If particles can escape at some other boundary after some time $t_{\rm esc}$ through, e.g., advection or diffusive escape, the average spectrum over the acceleration zone would reach a steady-state spectrum given by $\mathcal E^{\rm int}_p(t_{\rm esc})$. If $t_{\rm esc}\lesssim t_{\rm d}$, this spectrum would scale as $\propto p$ for $p<p_{\rm pk}(t)$, see Eq.~(\ref{eq:Eplin}), while if $t_{\rm d}\lesssim t_{\rm esc}\lesssim t_\times$, damping would lead to the broken power-law spectrum characterized by Eq.~(\ref{eq:Epnonlin1}) and (\ref{eq:Epnonlin2}), {\it viz.} hard below $p_{\rm pk}$, flat ($s\sim2$) above, up to the momentum $p_{\rm esc}=p_{\rm d}(t_{\rm esc})$, beyond which it would turn over into the high-energy tail of the spectrum accelerated without damping, as before. Interestingly, that part of the spectrum declines more slowly in momentum than a naive exponential cutoff. 
Finally, if $t_{\rm esc}>t_\times$, one should include the segment $\propto  p^{0.5-0.7}$, as before.

Considering now energy losses characterized by a momentum $p_{\rm loss}$, defined as that where the energy loss rate outpaces that of energy gain, one can anticipate the following. At times $t$ such that $p_{\rm d}(t)\ll p_{\rm loss}$, the time-dependent spectra derived in Sec.~\ref{sec:impl} would remain unchanged. Once $p_{\rm d}(t)\sim p_{\rm loss}$, the time-dependent spectra should freeze with $p_{\rm d}(t)\sim p_{\rm loss}$ at all subsequent times. The magnetic energy that flows into the cascade down to $\hat k^{-1}\sim \hat p_{\rm loss}$ would then be directed into particles as before but reconverted about as fast into radiation. 
If $p_{\rm loss}\lesssim p_{\rm pk}$, turbulent damping can be neglected of course, hence the steady-state spectrum is expected to exhibit a pile-up shape around the critical momentum at which the acceleration and cooling balances, as is well-known in classical stochastic acceleration~\cite{1984A&A...136..227S}. For $p_{\rm loss}\gtrsim p_{\rm pk}$, however, this pile-up would become less prominent due to two reasons. First, the energy spectrum is flatter, ${\mathcal E}_p\propto p^0$ or ${\mathcal E}_p\propto p^{0.5-0.7}$, so the number of particles above the critical momentum gives a modest correction. Second, the acceleration essentially stops at higher energies due to the damping. 

\subsection{Some phenomenological consequences}
Overall, one may thus expect a generic spectrum of a broken power-law form, with a hard spectral slope below $p_{\rm pk}$,
and a slope $s\sim 2$ above. The spectral slope below $p_{\rm pk}$ is left unspecified here, because it depends on the degree to which residual acceleration can proceed around $p_{\rm pk}$ as well as on the injection rate of particles, as discussed above. 
If such residual acceleration can be neglected, the location of the break is given by Eq.~(\ref{eq:physdamp_p}); otherwise, $p_{\rm pk}$ may increase in time. Assuming that, at the initial time, the nonthermal particle energy density represents a fraction $x_{\rm nth}<1$ of the thermal plasma content, then $p_{\rm pk}\sim p_0\, \beta_{\rm A}^{1-q}\beta^{-1}x_{\rm nth}^{-1}$, with $p_0$ the characteristic thermal momentum. This break momentum can thus take large values depending on the parameters $\beta_{\rm A}$, $q$ and $x_{\rm nth}$. Interestingly, broken power-law spectra characterized by a hard low-energy segment, a high-energy flat ($s\sim 2$) spectrum and a break at large momenta, as above, often emerge in the phenomenological modeling of high energy sources such as blazars~\cite{1998ApJ...509..608T,2018Galax...6..116R,2020NatAs...4..124B}, gamma-ray bursts~\cite{2004RvMP...76.1143P} or pulsar wind nebulae~\cite{2019hepr.confE..33A}, see e.g. Refs.~\cite{2021PhRvD.104j3005Z,2022MNRAS.517.2502S,2023ApJ...953..116L} for recent discussions in the present context. The present findings thus open interesting avenues of research for phenomenological applications in high-energy multi-messenger astrophysics.

The Crab nebula, for instance, represents a prime candidate for the present scenario because the characteristic eddy turn-over timescale can be as short as a year, $t_{\rm nl}\sim 1\,{\rm yr}\,(\ell_{\rm c}/0.3\,{\rm pc})$ assuming $v_{\rm A}\sim c$, given that the coherence length cannot indeed exceed the size of the nebula which is of the order of a pc, while the pulsar has been injecting energy into the nebula for about a thousand years. In view of the rather large characteristic magnetization $\sigma\simeq (v_{\rm A}/c)^2 \lesssim 1$ that is inferred from dynamical arguments~\cite{1984ApJ...283..694K,2009ASSL..357..421K}, numerical simulations~\cite{2014MNRAS.438..278P}, and phenomenological broad-band modeling~\cite{2023A&A...671A..67D}, the conditions for back-reaction should be amply satisfied. Interestingly, the spectral energy distribution can be generally well reproduced by a model in which the electrons and positrons are distributed as a broken power-law in momentum characterized by a break at Lorentz factor $\gamma_{\rm bk}\sim 2\times 10^5$, and spectral slopes $s_1\sim 1.5$ below the break, $s_2\sim 3.1$ above the break. The latter actually corresponds to an accelerated population with slope $s_{\rm acc,2}=s_2-1\sim 2.1$ because synchrotron radiation in a magnetic field of strength $B\sim 300\,\mu$G effectively cools electrons down to a Lorentz factor $\gamma_{\rm syn} \sim 3 \times 10^5$ over the pulsar lifetime, i.e., in a marginally fast cooling regime. In the frame of the present discussion, we speculate that the pairs could be injected in a relativistic turbulent bath in the central parts of the nebula, where the broken power-law spectrum is shaped quasi-instantaneously by stochastic acceleration, until damping freezes the spectrum. The pairs would then cool as they diffuse outwards in the nebula. At magnetization $\sigma \sim \mathcal O(1)$, Eq.~(\ref{eq:physdamp_p}) suggests that $p_{\rm pk}\sim p_0$, indicating that the break Lorentz factor would correspond to that at which particles are injected into the nebula. The arguments developed above (Sec.~\ref{sec:loss}) suggest that this picture could potentially reproduce the general features of the particle energy distribution. Assuredly, more work is needed to put this model on a firmer footing, in particular for what concerns the origin of the so-called radio electrons~\cite{2019hepr.confE..33A}.

In Sec.~\ref{sec:impl}, we have also observed that the spectrum can turn over into a hard spectral shape with $s\sim 1.3-1.5$ at the highest energies. Such a regime would be of particular interest for theoretical models of ultrahigh-energy cosmic ray origin, because it produces a hard spectrum close to the confinement scale where $r_{\rm g}\sim \ell_{\rm c}$, {\it i.e.} close to the Hillas limit, as phenomenologically inferred from fits to the observed spectrum and composition~\cite{2023JCAP...05..024A}. 
Furthermore, the overall spectral shape guarantees that a substantial fraction, if not most of the particle energy, resides at the highest energies, thus alleviating somewhat the constraint on the overall energy budget of the source. 

In turbulent coronae of black hole accretion systems such as AGN, our results have new implications for the maximum momentum of cosmic-ray ions. If ${\mathcal E}_p$ is sufficiently small at the injection momentum, i.e., only a fraction of the particles is injected, the spectrum can be hard with $s\sim 1$ at energies of interest. Classically, with radiative losses, it can be harder due to the pile-up effect followed by a cutoff around $\sim p_{\rm loss}$ (e.g., Fig.~S3 of Ref.~\cite{2020PhRvL.125a1101M}). 
However, instead, the peak momentum could be set by Eq.~(\ref{eq:physdamp_p}). If ${\mathcal E}_p$ is sufficiently large at the injection energy, our phenomenological model predicts that the energy spectrum has a flat portion above the injection momentum, followed by a cutoff set by $\sim p_{\rm loss}$. 
In either case, particles can be injected through magnetic reconnections in the high magnetization region, where their small volume fraction could lead to the small number of injected particles. The cosmic-ray pressure could be saturated to be a decent fraction of the magnetic pressure, due to damping.     

Finally, one obvious consequence of turbulence damping is to increase the polarization degree of the emitting zone, possibly up to maximal values if the emitting zone is restricted to one coherence length volume and if damping is complete up to scale $\sim \ell_{\rm c}$.

\section{Summary and conclusions}
Building up on earlier work, this paper has provided a comprehensive study of nonlinear stochastic acceleration, a regime in which the accelerated particles draw sufficient amounts of energy to alter the energy cascade of the turbulence, which itself regulates the efficiency of the acceleration process. To model this nonlinear coupling between particles and turbulence, we have proposed and analyzed a general phenomenological model that solves a transport equation in momentum space for the particles, the cascade equation in wavenumber space for the turbulence, accounting for a sink of energy due to particle acceleration, and accounting altogether for the dependence of the cascade and acceleration rates on the amount of energy in each channel. We have then determined the generic spectral energy distributions for the particles and the turbulence as a function of ambient physical conditions, and time. 

The history of the overall process can be summarized as follows. As particles draw energy from the turbulence, their spectra shift to larger and larger momenta, or equivalently larger and larger length scales if momenta are expressed as gyroradii. An important specificity of stochastic acceleration is its tendency to shape hard energy spectra in the stationary limit, which means that as time goes on, particles draw more and more energy from the turbulent cascade. In detail, if the momentum diffusion coefficient $D_{pp}\propto p^r$, the steady-state energy density spectrum (per log-momentum interval) evolves as $\mathcal E_p\propto p^{3-r}$ in the absence of turbulent damping and nonlinear feedback. This corresponds to a particle momentum distribution ${\rm d}n/{\rm d}p\propto p^{-2}\mathcal E_p\propto p^{1-r}\propto p^{-s}$, with $s\simeq r-1$, where $n$ represents the particle density.

In the stochastic Fermi process, particles of a given gyroradius $r_{\rm g}$ draw energy from a certain range of length scales $l$, generally $l\gtrsim r_{\rm g}$. Provided the turbulence is externally stirred over sufficiently long timescales, there exists a time $t_{\rm d}$ at which particles of momentum $p$ start to draw energy from the turbulence at a rate comparable to that at which turbulent energy flows through the cascade on the relevant length scale $l$. Specifically, this momentum $p=p_{\rm pk}$ corresponds to the momentum at which $\mathcal E_p$ peaks at $t_{\rm d}$ and the length scale $l\sim r_{\rm g}(p_{\rm pk})$. Once that point is reached, turbulent damping becomes significant. This, in turn, tends to suppress acceleration on length scales $\lesssim l$, and therefore to freeze the particle spectra below  $p_{\rm pk}$. Meanwhile, particles of larger momenta keep being accelerated and concomittantly damp the turbulence on increasing length scales. At any given time beyond $t_{\rm d}$, the spectrum is thus frozen below a momentum $p_{\rm d}(t)$, which keeps increasing in time, because the turbulence has undergone damping below length scales $\sim r_{\rm g}(p_{\rm d})$. Overall, this shapes a broken power-law spectrum, with a hard component below $p_{\rm pk}$, and a flat spectrum ${\rm d}n/{\rm d}p\propto p^{-s}$ with $s\sim 2$, from $p_{\rm pk}$ up to $p_{\rm d}(t)$. Beyond $p_{\rm d}(t)$, the spectrum recovers the form it would have in the absence of damping, see Fig.~\ref{fig:fdbck} for an illustration at different times.

As time goes on, the particle energy spectrum extends in momentum space (hence to larger length scales in terms of gyroradius), just as the (time-dependent) maximum length scale $r_{\rm g}[p_{\rm d}(t)]$ below which the turbulent cascade is damped. Because the energy content of the cascade increases with length scale, it may be that the particle energy density at that scale eventually becomes comparable to the cascade energy density at wavenumber $k\sim r_{\rm g}[p_{\rm d}(t)]^{-1}$. If this point is reached, the particle energy spectrum changes slope, generating a hard segment with $s\sim 1.3-1.5$.  However, back-reaction becomes effective only once the particle energy density crosses a certain threshold, see  Eq.~(\ref{eq:physdamp}). In practice, that constraint implies that the segment with $s\sim 1.3-1.5$ should be limited to a restricted dynamic range of gyroscales close to the outer scale of the turbulence.

In summary, one should thus observe the following. On timescales $t\lesssim t_{\rm d}$, damping and nonlinear feedback can be ignored. This timescale $t_{\rm d}$ can be derived from Eqs.~(\ref{eq:physdamp}) or Eq.~(\ref{eq:physdamp_p}). On longer times, $t\gtrsim t_{\rm d}$, damping remodels the spectrum into the broken power-law form discussed above, namely a hard spectrum at $p<p_{\rm pk}$, a flat ($s\sim 2$) spectrum above, including the fall-off above $p_{\rm d}(t)$, and possibly the segment $s\sim 1.3-1.5$ at the highest energies. In this description, the break momentum $p_{\rm pk}$ remains frozen in time. These spectra assume impulsive injection of particles at the initial time at momentum $p_0$ and thus represent Green's functions. The time-integrated spectra with continuous particle injection preserve the above features, except for the low-energy part below $p_{\rm pk}$, which takes the form of a power-law with index $s\sim 1$ in the absence of damping (assuming $D_{pp}\propto p^2$), or harder once damping sets in.

As we have stressed, the present discussion remains exploratory, given the numerous uncertainties that plague the nature of particle acceleration in magnetized turbulence. We have discussed a number of limitations and nevertheless conclude that the above picture appears globally robust. It seems likely that some form of secondary acceleration could develop below the momentum $p_{\rm pk}$ at which back-reaction first sets in. This would imply that the spectrum in that region is not frozen in time but would evolve, likely at a slower rate than the high-energy extension characterized by the slope $s\sim 2$. One direct consequence is to shape a hard spectrum below that critical momentum, whose exact slope would be dictated by the rate of acceleration in that region and the rate at which particles are injected in the process. We have also briefly addressed the influence of escape and losses on the general shape of the spectrum.

As we have noted, broken power-law spectra of the above form emerge as generic features of phenomenological models in high-energy multi-messenger astrophysics. Our results thus open interesting avenues of research for phenomenological applications, which we plan  
to address in forthcoming studies.


\medskip
\begin{acknowledgments}
The work of K.M. was supported by the NSF Grants No.~AST-2108466 and No.~AST-2108467, and KAKENHI No.~20H01901 and No.~20H05852. F.R. acknowledges support by a DFG grant RI
1187/8-1.
The work of M.L. has been supported by the ANR (UnRIP project, Grant No.~ANR-20-CE30-0030). M.L. and K.M. warmly acknowledge the hospitality of the Institute for Advanced Study (Princeton) and of the Department of Astrophysical Sciences of Princeton University, where this work was initiated. 
\end{acknowledgments}


\appendix

\section{Numerical parameters}\label{sec:appA}
We have considered two generic kernels: one describing gyroresonance, the other describing nonresonant interactions with larger-scale modes. The first kernel is written as a Gaussian in log-space
\begin{equation}
\Phi(k;\,p) \propto \exp\left[-\frac{\left(\ln \hat p + \ln \hat k\right)^2}{2 \Delta_1^2}\right]\quad \left({\rm gyroresonant}\right)\,,
\label{eq:phik1}
\end{equation}
with $\Delta_1=1.0$, whose value is purely ad hoc. 

\begin{figure}[h]
\includegraphics[width=0.48\textwidth]{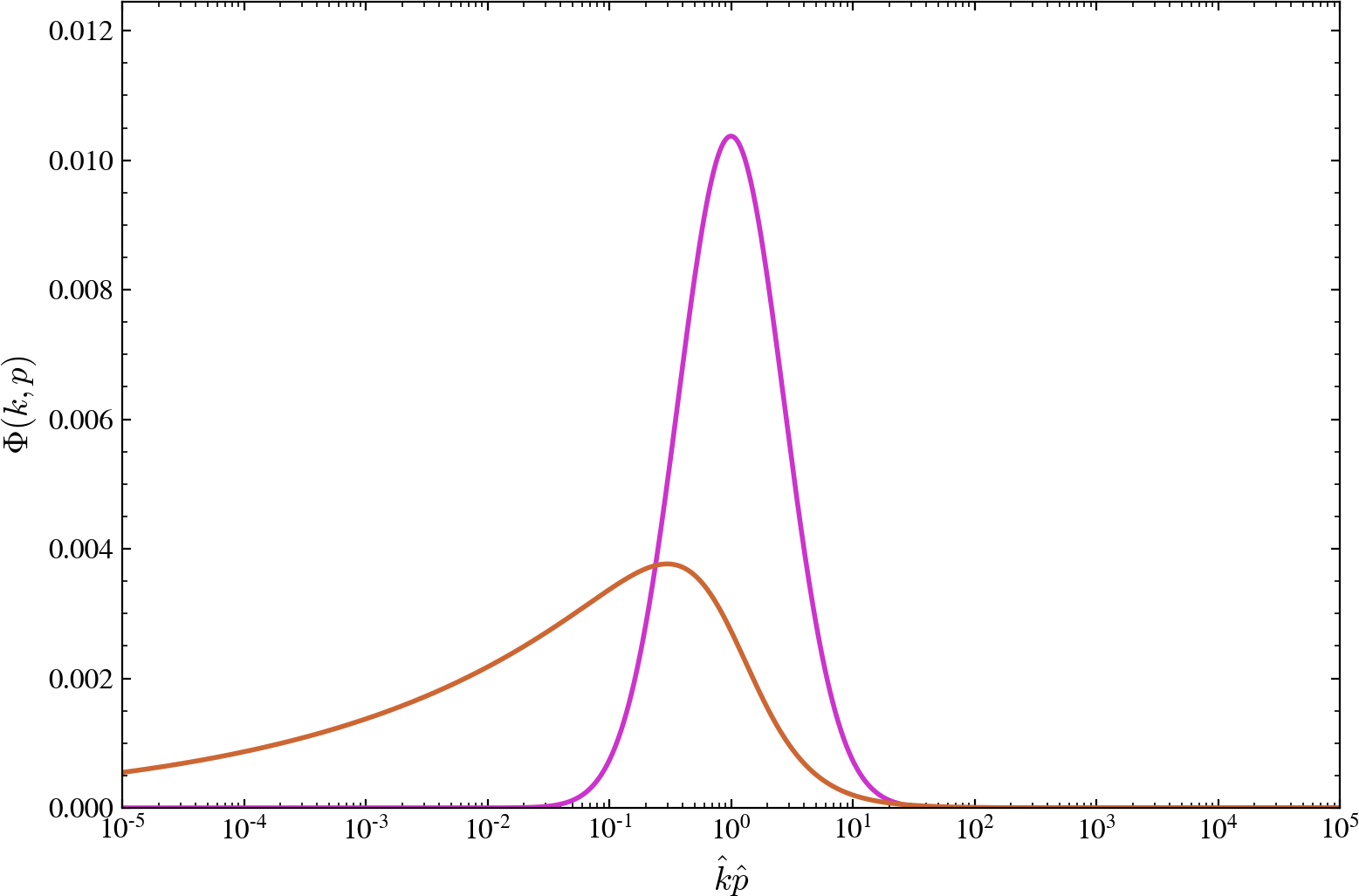}
\caption{Kernels $\Phi(k;\,p)$ defined in Eq.~(\ref{eq:phik1}) and (\ref{eq:phik2}) describing gyroresonant (purple) and nonresonant (orange) interactions, respectively. Those kernels determine how the energy gained by particles of momentum $p$ ($\hat p\equiv p/p_0$) is distributed as loss in the turbulence cascade in terms of wavenumber $k$ [$\hat k \equiv k\,r_{\rm g}(p_0)$]. 
\label{fig:Phi}}
\end{figure}

\begin{figure}[t]
\includegraphics[width=0.48\textwidth]{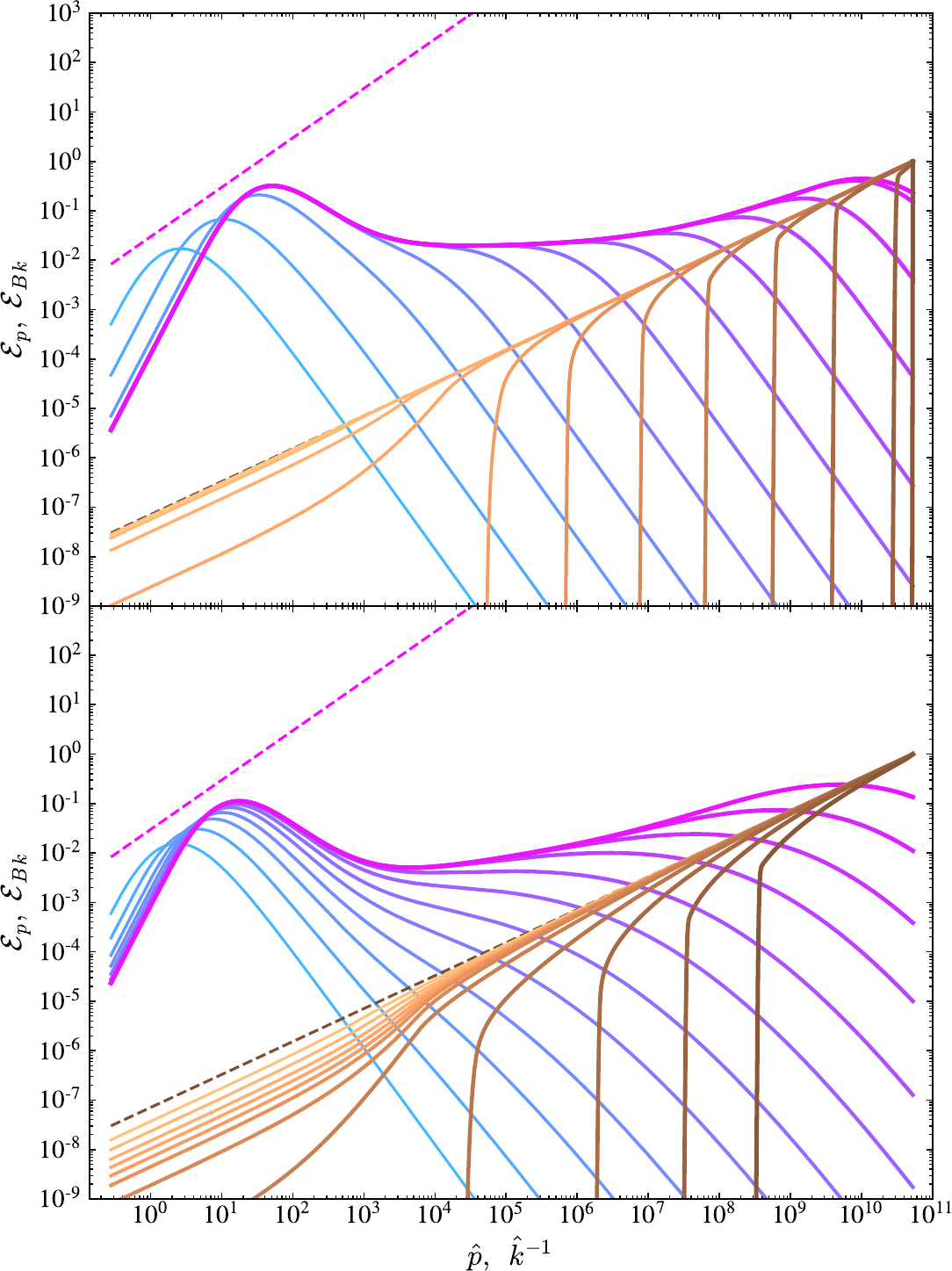}
\caption{Comparison of the results for a gyroresonant kernel (top panel) and a nonresonant kernel (bottom panel) as defined in the text and displayed in Fig.~\ref{fig:Phi}. The other parameters used to the numerical integration are those corresponding to the top panel of Fig.~\ref{fig:var}.
\label{fig:varPhi}}
\end{figure}

The second kernel is biased toward larger scales, and we select in particular the following broken power-law shape
\begin{align}
    \Phi(k;\,p) \propto &\left\{\exp\left[\ln\left(\hat p\hat k\right)/\Delta_2\right]+\exp\left[-\ln\left(\hat p\hat k\right)/\Delta_3\right]\right\}^{-1}\nonumber\\
    &\quad\quad \left({\rm nonresonant}\right)\,,
\label{eq:phik2}
\end{align}
where $\Delta_2=0.7$ and $\Delta_3=5.0$. Those two kernels are illustrated in Fig.~\ref{fig:Phi} and the consequences of using one or the other are shown in Fig.~\ref{fig:varPhi}, for the numerical parameters corresponding to the case shown in the top panel of Fig.~\ref{fig:var}. Although the kernel parameters are ad hoc, they suffice for the purpose of demonstrating the impacts of the non-resonant interaction.  Dedicated numerical simulations might hopefully shed light on these parameters. In the main text, we discuss only the results using the first (gyroresonant) kernel only.

Regarding the numerical scheme used to integrate the model equations, we rely on an implicit Crank-Nicholson method, using a high-order stencil for the spatial derivatives. The large dynamic range in $k$ and $p$ renders the system unstable: advection rate in wavenumber space increase with $k$, therefore decrease with $k^{-1}$, while the energization rates rather increase with $p$. It has proven necessary to introduce some regularization procedures, in particular a minimal floor value for $\gamma_k$ (where the turbulence has been damped) and to smooth the numerical solutions at each time step with a Gaussian filter of one bin width, in order to quench numerical instabilities that grow on the mesh size. The dynamic range has here been set to $2\times 10^{11}$. In Fig.~\ref{fig:fdbck}, turbulence is driven at the outer scale by maintaining the value of ${\mathcal E_B}_{k_{\rm inj}}$ constant, which corresponds to the value at the maximum of ${\mathcal E_B}_k$. The loss of particles through both boundaries is minimal, as is obvious from Fig.~\ref{fig:fdbck}. Similar spectral shapes would be obtained by changing the dynamical range, provided it remains larger than a few decades, as we have checked. Finally, with regards the choice of numerical parameters, we have integrated the system for various values of $\Delta_0$, $\Delta_1$ and $\Delta_2$ that control the width of $\Phi(k;\,p)$. The observed spectra were similar in all cases.


\bibliography{kmurase.bib}

\end{document}